\documentclass[journal]{IEEEtran}

\usepackage{cite}

\ifCLASSINFOpdf
   \usepackage[pdftex]{graphicx}

\else

\fi

\usepackage[cmex10]{amsmath}
\usepackage{amsfonts}
\usepackage{amsthm}

\usepackage{algorithm}
\usepackage{algpseudocode} 
\usepackage{subfigure}
\usepackage[utf8]{inputenc}
\usepackage[english]{babel}

\usepackage{bm}
\usepackage{color}

\begin{document} 

\title{Environment-Aware Minimum-Cost Wireless Backhaul Network Planning with Full-Duplex Links}
		\author{Omid~Taghizadeh,~\IEEEmembership{Student Member,~IEEE},~Praveen~Sirvi,~Santosh~Narasimha,~Jose~Angel~Leon~Calvo,~\IEEEmembership{Student Member,~IEEE},~Rudolf Mathar,~\IEEEmembership{Senior~Member,~IEEE}
			\IEEEcompsocitemizethanks{
		\IEEEcompsocthanksitem Authors are with the Institute for Theoretical Information Technology, RWTH Aachen University, Aachen, 52074, Germany (email:  \{taghizadeh,sirvi,narasimha,leon,mathar\}@ti.rwth-aachen.de).
					%\IEEEcompsocthanksitem A.~C.~Cirik and L. Lampe are with the Department of Electrical and Computer Engineering, University of British Columbia, Vancouver, BC V6T 1Z4, Canada  (email: \{cirik,~lampe\}@ece.ubc.ca).
					\IEEEcompsocthanksitem Part of this paper is accepted for presentation at the 2018 IEEE Wireless Communications and Networking Conference (WCNC'18) \cite{Tagh1804:Minimusm}. 
%\IEEEcompsocthanksitem O.~Taghizadeh,  are with the Institute for Theoretical Information Technology, RWTH Aachen University, Aachen, 52074, Germany (email: {taghizadeh, mathar}@ti.rwth-aachen.de). 
							}}  

% make the title area
\maketitle

\begin{abstract}
In this work, we address the joint design of the wireless backhauling network topology as well as the frequency/power allocation on the wireless links, where nodes are capable of full-duplex (FD) operation. The proposed joint design enables the coexistence of multiple wireless links at the same channel, resulting in an enhanced spectral efficiency. Moreover, it enables the usage of FD capability when/where it is gainful. In this regard, a mixed-integer-linear-program (MILP) is proposed, aiming at a minimum cost design for the wireless backhaul network, considering the required rate demand at each base station. Moreover, a re-tunning algorithm is proposed which reacts to the slight changes in the network condition, e.g., channel attenuation or rate demand, by adjusting the transmit power at the wireless links. In this regard, a successive inner approximation (SIA)-based design is proposed, where in each step a convex sub-problem is solved. Numerical simulations show a reduction in the overall network cost via the utilization of the proposed designs, thanks to the coexistence of multiple wireless links on the same channel due to the FD capability.      
\end{abstract}


\IEEEpeerreviewmaketitle

\section{Introduction} \label{sec_intro}

Deployment of small cell base stations (SCBS) is a necessary paradigm to meet the rapid increase of rate demand in the context of the fifth generation cellular wireless networks (5G). In particular, small cells enable high capacity radio access solutions due to the short distance coverage, facilitating a higher energy and spectral efficiency~\cite{5723093}. However, with the expected dense deployments of SCBS, the main challenge for the operators will be to provide backhaul solutions at a reasonable cost, in order to handle the data traffic to (from) the core network. In this regard, application of wireless backhaul solutions appears to be a good alternative to the traditional fiber connections, due to the lower capital expenditure and ease of deployment~\cite{6963798, 6544262}. However, wireless backhaul links are known to suffer from occasional failure/degradation due to blockage and weather conditions, limited information capacity compared to the traditional fiber connections, as well as the additional consumption of energy and spectrum. In order to overcome the aforementioned drawbacks, several works have been dedicated to the efficient design of the wireless backhaul networks from the aspects of link/topology planning and resource allocation \cite{6608635, charnsripinyo2003topological, 978050, 5683870, monti2012mobile, 6133999, 7194106, 6876516, li2016multi,nadiv2010wireless, 1191628}, as well as exploring promising technologies with the goal of enhancing the performance of the traditional line-of-sight (LOS) point-to-point microwave links. This includes, e.g., establishment of non-LOS links for resolving blockage situations \cite{coldrey2012small, coldrey2013non}, or technologies with the potential to enhance spectrum utilization, e.g., operating in unlicensed millimeter-wave bands \cite{hur2013millimeter}, optical links \cite{li2015optimization}, as well as the realization of in-band backhauling, i.e., the co-existence of access and backhaul links at the same channel~\cite{7306541, 7177124, rahmati2016price, KRSV:16, 7247153, AK:15}.

Among the promising technologies for the efficient usage of spectrum, full duplex (FD) capability is introduced as the transceiver's capability to transmit and receive at the same time and frequency, however, resulting in a strong self-interference~\cite{7219383, 7393479}. Recently, effective methods for self-interference cancellation (SIC) are proposed \cite{Bharadia:14,YLMAC:11,BMK:13,khandani2013two}, motivating further studies on wide range of related applications \cite{HBCJMKL:14,SSDBRW:14}. It is shown in \cite{CTLMH:16, ZTH:131, ZTH:132, DBLP:journals/corr/TaghizadehRCML17, taghizadeh2017linear, Tagh1710:Power, Tagh1705:Sum, 7997171, 8234646, TM:14:1, Taghizadeh2016} that the application of FD capability improves the spectral efficiency of the point-to-point or multi-hop wireless communications, relying on an optimized resource allocation and power control. In particular, the utilization of the FD capability at wireless backhaul links is presented as a promising use case, due to the zero-mobility conditions and the utilization of directive antennas, hence showing a potential to reduce the cost of spectrum~\cite{HBCJMKL:14}.   

\subsection{Related works} 
\subsubsection{Wireless backhaul planning}
The efficient planning of a wireless access network has been the focus of several recent works~\cite{li2016multi}. In this regard, the first step is to determine the location and the required number of SCBS sites, considering the available resources from the operator, distribution of users, as well as the expected rate demands~\cite{1226011, juttner2005two, wu2005optimization}. A proximity-based clustering method is proposed in ~\cite{1191628}, associating the groups of given SCBS sites to the available root node, i.e., a gateway node to the core network via fiber connection. Once the clusters of SCBS are identified in association with the root nodes, the remaining task is to optimally plan the backhauling network within each cluster. This includes the choice of the links to be established, identifying the network topology, as well as the link specifications, e.g., link technology and the operating frequency bands. In this regard, the problem of backhaul topology planning and optimization has been addressed with the considerations of energy efficiency \cite{6963798, monti2012mobile, 6133999}, overall network cost \cite{7194106, 6876516, li2016multi,nadiv2010wireless,5683870}, delay performance \cite{7374694,7165594}, as well as reliability and fault tolerance \cite{6608635, charnsripinyo2003topological, 978050, 5683870}. The problem of resource allocation and frequency planning for a network with a fixed topology is studied in \cite{5743593, yi2012backhaul}. In the recent works \cite{6876516, li2016multi} a multi-objective design is considered, accounting for the impact of interference, delay, and system throughput. However, the aforementioned works are not yet extended for a joint frequency/topology wireless backhaul planning, which is essential for the scenarios with the high impact of interference, e.g., dense urban deployment.  

%The objective of the wireless backhaul planning is to specify the location of backhaul sites, as well as the choice of the wireless links to be established, as well as the operating frequency bands for each link. In this respect, the first step is to specify the location of small cell nodes, as well as the SCBS nodes to be served. In this regard, a proximity based clustering method is proposed in \cite{}, associating the SCBS sites to an optimal root node location. 
%
%Once the location of the backhaul sites are specified, the second step is to perform a topology/link planning, given the backhaul site locations. 
\subsubsection{FD-enabled communication} 
In \cite{AK:15} the performance of an FD in-band backhauling system is analyzed, by means of stochastic geometry, where the superior performance of an FD-enabled system is observed compared to the half-duplex (HD) counterparts. An adaptive FD/HD backhauling system is studied in \cite{7247153} where FD in-band backhauling is used in a two-tier star topology. The aforementioned system is also extended with the considerations of system sum-rate analysis and optimization \cite{KRSV:16}, minimum-cost resource allocation \cite{rahmati2016price}. In \cite{7306541, 7177124} a flexible frame structure is used to jointly optimize the access and backhaul parameters. In all of the aforementioned works, the FD capability is considered on a single backhaul-access link, or on a fixed star topology.  % Please note that the coexistence of the multiple wireless links at the same channel adds to the significance of an optimized network topology/frequency planning. However, 
%%%To the best of the authors knowledge, the topology/frequency planning for an FD-enabled backhauling network is not addressed in the available literature.                       

%-- wireless backhaul planning
%-- FD assisted backhauling

\subsection{Contribution}
In this paper we address the joint topology design and resource allocation in a wireless backhaul network where FD capability is enabled at the wireless links, with the goal of minimizing the collective network cost. The contributions of this paper are summarized as follows:
\begin{itemize}
\item In contrast to \cite{7306541, 7177124, rahmati2016price, KRSV:16, 7247153} where FD capability is studied for a single backhaul link or a fixed star topology, we consider a general topology with the co-existence of multiple backhaul and access links on the same channel. However, this calls for an efficient interference management scheme. In order to facilitate this, we consider a framework where the interference conditions for backhaul-to-access, as well as the backhaul-backhaul links can be obtained via the utilization of a wave propagation simulator, suited for dense urban scenarios where relatively accurate environment information is available~\cite{ScReMa11,ScReMa12}. 
%This is in contrast to the designs in \cite{7194106, 5683870} assuming no interference from backhaul links, or the recent works \cite{6876516, li2016multi}, that simplify the impact of interference to link crossing and low arrival angle situations.  
\item Unlike the connected graph approaches proposed in \cite{7194106, 6876516, li2016multi, nadiv2010wireless, 5683870} or the frequency planning on a fixed topology \cite{5743593, yi2012backhaul}, we jointly address the design of the network topology as well as the allocation of the power and frequency resources, as the aforementioned factors jointly impact the network interference pattern and the resulting link throughput. In particular, this enables a flexible usage of FD capability, when and where it is gainful. In this regard, an {MILP} design framework is proposed to obtain a minimum cost network operation, complying with required QoS as well as the operational network constraints. 
\item Due to the modifying nature of the network data, e.g., change of traffic load or wireless link conditions, the networks are usually adjusted to the worst-case conditions, resulting in a reduced efficiency. In this regard, we propose an {SIA} design framework for adjusting the transmit power at the wireless links, with the goal of keeping the network in compliance with the changing channel or QoS conditions. In contrast to the introduced proactive measures in \cite{6608635, charnsripinyo2003topological, 978050, 5683870}, where the network is designed with the consideration of a possible failure, our proposed approach is a reactive one; enabling the network to react to the partial changes with a minimal cost.   
\end{itemize}
Numerical simulations show a reduction in the network cost via the utilization of the proposed designs, thanks to the coexistence of multiple links on the same channel due to the FD capability. 
%\subsection{Notation}
%
%

\section{Network model}\label{sec_model}
\begin{figure}[!t] 
    \begin{center}
        \includegraphics[angle=0,width=0.99\columnwidth]{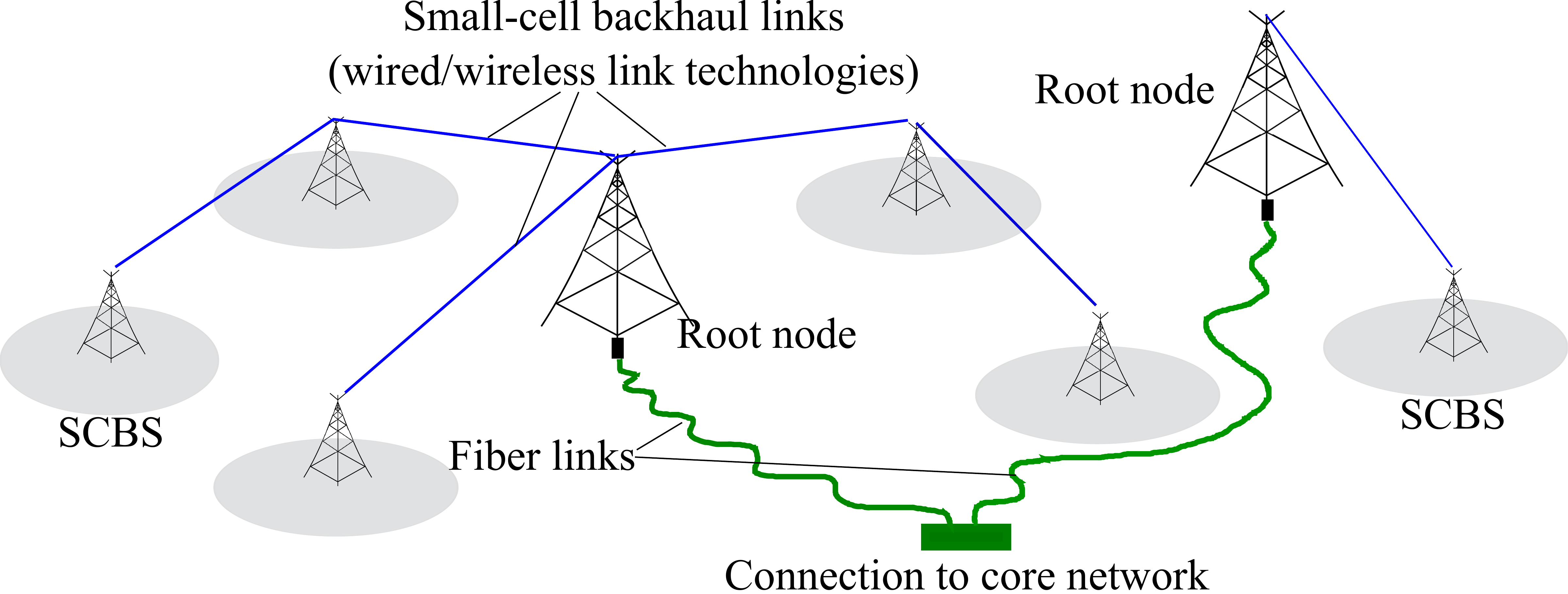}
				        %\fbox{model_rect6.pdf}
    \caption{Backhauling network, including wireless and wired link technologies for small cells and fiber connections for root nodes to the core network. } \label{FD_Cellular_fig_backh_net_system}
    \end{center} \vspace{-0mm} 
\end{figure} 
We consider a network of backhaul nodes, including root nodes and non-root nodes, whereby the data traffic is carried to (from) the access network from (to) the core network, see Fig.~\ref{FD_Cellular_fig_backh_net_system}. Root nodes are connected to the core network via high capacity fiber connections, whereas non-root nodes set up wireless links directed to the root nodes, or to the other non-root nodes which relay their data traffic through the network. The goal of the network is hence to deliver the required uplink (downlink) information rate from (to) the non-root nodes to (from) the core network, while complying with the operational network constraints. For each wireless link the transmit power can be adjusted for different frequency subchannels, thereby facilitating an effective interference control and spectrum-saving mechanism. The wireless channels associated with each link, as well as the interference channels between different links are assumed to be frequency-flat at each sub-channel. 
\subsection{Acquisition of network information} \label{FD_Cellular_model_net_info}
In order to deliver a mathematical description of the network operation, the following information is obtained via direct observations and measurements, or via post-processing of the observable data:   
%\begin{itemize}
\subsubsection{Node locations} The topology of the network is partly determined by the location of the root and non-root nodes. Moreover, based on the location of the nodes, the potential wireless links are determined. This can be identified via the existence of a LOS between two nodes, and that the corresponding link does not exceed a maximum distance limit \cite{6876516, li2016multi}. The set of root nodes, non-root nodes, and all nodes are respectively denoted as $\mathbb{R}, \mathbb{M}$ and $\mathbb{N}$. The set of all potential links is denoted as $\mathbb{L}$. 
\subsubsection{Available spectrum} The information regarding the provisioned operational spectrum, including the spectrum dedicated for backhaul and access are obtained. The set of all frequency subchannels, and those used in the access network are denoted as $\mathbb{F},\mathbb{F}_a$, respectively. 
  
\subsubsection{Large-scale channel parameters} The large-scale channel parameters associated with the desired and interference paths are obtained for each potential wireless link, and used as the basis for network planning. This includes the estimated channel strength, i.e., ratio between the transmit and receive power at each link. Note that the exact channel information may not be obtained via direct measurements, as all of the potential links may not exist prior to the network planning and realization. However, the large-scale channel conditions can be estimated at each link via the application of a wave propagation simulator \cite{ScReMa12} utilizing the used antenna specifications as well as the related environment information, e.g., city map and node locations. It is worth mentioning that for the studied system this process reaches a high accuracy, considering the static wireless links with LOS connections and almost zero mobility, which reduces the randomness. The interference channel strength between a link and the access network, e.g., from a wireless link to a remote BS, can be obtained by following a similar methodology. The desired channel strength associated with the wireless backhaul link $(i,j)\in \mathbb{L}$ and the interference channel strength from the wireless link $(l,k)\in \mathbb{L}$ to the link $(i,j)$ are denoted as $\Lambda_{ij,f}$ and $\Gamma_{ij,lk,f}$, respectively. Moreover, the interference channel strength from the wireless link $(i,j)$ to the access network associated with node $m$ is denoted as $\Omega_{{ij},m,f}$, where $f\in\mathbb{F}$ and  $i, j, m \in \mathbb{N}$.     %\cite{xxx} for a detailed description of the related methodologies.%% Expl: once a netwoek is setup, desired and interfernece channels are obtained via measurements, to retune the network.  
% Expl: SINR s are given from accurate channels, argue that they are accurate.
\subsubsection{SIC level} Due to the full-duplex capability, a node may transmit and receive via separate wireless links at the same frequency, however, resulting in a strong SI. Such interference can be reduced via the application of state of the art SIC schemes, e.g., \cite{Bharadia:14,BMK:13}. As mentioned, a wireless backhaul link is particularly interesting for the utilization of FD capability due to the almost zero-mobility condition and the high cost of spectrum. However, the SIC may not be perfect due to the impact of hardware inaccuracies, as well as the reflections from the moving objects. In this regard, the imperfect SIC can be modeled as the attenuation factor $\Gamma_{ij,ki,f}$, relating the transmission power to the resulting residual self-interference power at the node $i \in \mathbb{N}$ and from the link $(i,j)$ to $(k,i) \in \mathbb{L}$.    
\subsubsection{Average processing delay} The processing delay at the intermediate nodes, denoted as $d_i, \; i \in \mathbb{N}$, plays a dominant role in the overall delay for the transfer of information from (to) the core network~\cite{7374694}. Please note that the average processing delay can be considered as a function of the available processing and storage resources at each node, and is assumed to be a known information in our work.     
\subsubsection{Power budget} The maximum transmit power at each wireless link, denoted as $P_{\text{max},ij}$ as well as the total consumed power at each node, denoted as $P_{\text{max},i}$, should be considered in the design of network parameters. Note that the value of $P_{\text{max},ij}$ is usually limited by the range of the transmit chain elements, e.g., power amplifier, whereas the $P_{\text{max},i}$ is limited by the available power sources, e.g., when the node is battery-powered or relies on the harvesting sources.  
\subsubsection{Required information rate} The required information rate at each node is denoted as $R_{\text{ul},i}$, representing the UL traffic requirement, and as $R_{\text{dl},i}$, as the DL traffic requirement, $i \in \mathbb{N}$. The values of $R_{\text{ul},i}$, $R_{\text{dl},i}$ can be estimated considering the number of users associated with a backhaul node, or by learning the previous network demands. 
\subsubsection{Pre-existing links} Other than the wireless backhaul links, the network may make use of the pre-existing wired links, e.g, pre-existing cooper connections. The available collective link capacity from the other technologies is denoted as $C_{ij,0}$, for the link $(i,j) \in \mathbb{L}$. 
\subsubsection{Interference temperature threshold} Interference temperature threshold, denoted as $I_{\text{th},k,f}$, gives a trade-off between the protection of the access network against the interference from backhaul wireless links, and the coexistence possibility for backhaul and access links at the same frequency. For instance, it can be chosen equal to the the noise variance, i.e., keeping the interference below the noise floor. 
%\end{itemize}
\subsection{Wireless link throughput}
Once a wireless link is established, the link quality in terms of signal to interference-plus-noise ratio (SINR) is obtained as
\begin{align} \label{FD_Cellular_model_SINR_per_link}
\gamma_{ij,f} = \frac{ \Lambda_{ij,f} X_{ij,f} }{ \sum_{(l,k) \in \left\{ \mathbb{L} \setminus (i,j) \right\} } \Gamma_{ij,lk,f} X_{lk,f} + W_{ij,f}}, 
\end{align} 
where $X_{ij,f}$ and $W_{ij,f}$ respectively represent the transmit power and the thermal noise variance for the link $(i,j) \in \mathbb{L}$ at the subchannel $f \in \mathbb{F}$. $\gamma_{ij,f}$ is the SINR, resulting in 
\begin{align} \label{FD_Cellular_model_capacity_per_link}
C_{ij,f} = B \log_2 \left(1 + \gamma_{ij,f}\right),
\end{align} 
where $B$ is the bandwidth of each frequency subchannel, and $C_{ij,f}$ is the achievable information rate for the wireless link $(i,j)$ at the subchannel $f$. Note that the equality (\ref{FD_Cellular_model_capacity_per_link}) holds only for a Gaussian transmit signaling and for an arbitrarily long coding block length, and can be otherwise treated as an approximation. 

\subsection{Role of planning} \label{FD_Cellular_model_role_planning}
\begin{figure*}[!t] 
    \begin{center}
        \includegraphics[angle=0,width=1.8\columnwidth]{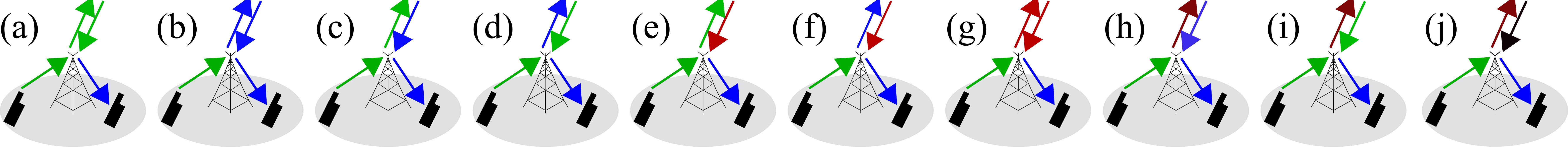}
				        %\fbox{model_rect6.pdf}
    \caption{Possible frequency duplexing modes for wireless backhaul connections. Different colors represent different frequency bands. This includes the possibility of multiple coexisting links at the same frequency, e.g., (a),~(b), or a complete separation of the links at different bands (j).} \label{FD_Cellular_interference_patterns}
    \end{center} \vspace{-0mm} 
\end{figure*}

As mentioned, each node may setup a wireless link among the identified feasible set $\mathbb{L}$, as a usual planning choice in the context of wireless backhaul planning. In this work, we also assume that the available spectrum is divided into multiple frequency subchannels, where the transmit power for each link can be adjusted at each frequency subchannel. The aforementioned flexibility enables the planning algorithm with two enhancements compared to the usual planning strategies \cite{7194106, 6876516, li2016multi,nadiv2010wireless,5683870}. Firstly, an effective interference management is enabled, regarding the interference to the other links, the SI in the FD mode, as well as the interference to the access network. In particular, it facilitates a proper incorporation of the FD capability by enabling different duplexing modes which are chosen depending on the interference conditions, SIC quality, and QoS requirements, see Fig.~\ref{FD_Cellular_interference_patterns}. Moreover, the required transmission rates can be obtained by adjusting the transmit powers, reducing un-necessary additional costs due to excessive spectrum and energy consumption.
Note that the expressions in (\ref{FD_Cellular_model_SINR_per_link})-(\ref{FD_Cellular_model_capacity_per_link}) formulate the achievable information rate of each wireless link, given the transmit strategies throughout the network. This is the main focus of the next part of this paper to identify how the aforementioned planning choices should be made in order to satisfy the network QoS requirements with a minimum overall costs.  

\section{Minimum-Cost Network Planning and Optimization}\label{sec_planning}
 %%%\begin{figure*}[!b]
%%%\normalsize
%%%{\small{\begin{align} \label{FD_Cellular_planning_link_capacity_Consts_2}
%%%C_{ij,f} \overset{(a)}{\leq}  & \underbrace{B \log_2 \left( \Lambda_{ij,f} X_{ij,f} + \hspace{-3mm} \sum_{(l,k) \in \left\{ \mathbb{L} \setminus (i,j) \right\} } \hspace{-3mm} \Gamma_{ij,lk,f} X_{lk,f} + W_{ij,f} \right) }_{=: f_1 (\mathbb{P})\geq   \mathcal{L}_p \left( \mathbb{P} ; \mathbb{P}_0 \right)}  - \underbrace{ B \log_2 \left(\sum_{(l,k) \in \left\{ \mathbb{L} \setminus (i,j) \right\} }\hspace{-5mm}  \Gamma_{ij,lk,f} X_{lk,f}  + W_{ij,f} \right)}_{\leq \mathcal{L}\left(  \mathbb{P} ; \mathbb{P}_0 \right)} \overset{(b)}{=: f_2 (\mathbb{P}) \geq} \mathcal{L}_p \left(  \mathbb{P} ; \mathbb{P}_0  \right) - \mathcal{L}\left(  \mathbb{P} ; \mathbb{P}_0 \right),
%%%\end{align} }}
%%%\hrulefill
%%%\vspace*{-0mm}
%%%\end{figure*}
%The goal of network optimization/planning is to satisfy the defined QoS, under the operational network constraints. 
In this part, two design strategies are proposed considering the defined wireless backhaul network. First, a full-scale network planning and optimization method is proposed, employing an MILP framework, where the planning choices explained in Subsection~\ref{FD_Cellular_model_role_planning} including the network topology, operational frequency bands, and the transmit power of wireless links are determined in order to setup a minimum-cost wireless backhaul network. However, this approach demands major re-configurations of the network parameters, e.g., topology, which may be only practical in the initialization phase of the network setup. Hence, as the second design strategy, the transmit power of the wireless links are adjusted in order to cope with slight changes in the network information, e.g., channel situation, SIC quality, or a slightly increased rate demand. The latter design acts as a network re-tunning method, assuming that the major network parameters including network topology and the functioning frequency bands are already fixed.

\subsection{Decision variables}
The following decision variables are considered as the outcome of our design, see Table~\ref{FD_Cellular_tab_dec_vars}:
\subsubsection{Transmit power $X_{ij,f}$} The decision variables $X_{ij,f}, \; (i,j) \in \mathbb{L}, \; f \in \mathbb{F}$, determine the transmit power of the wireless links, and at each frequency band.   
\subsubsection{Used link capacity for UL (DL) $C_{\text{ul},ij} (C_{\text{dl},ij})$} These variables determine the information flow within the network, regarding the used capacity of each link $(i,j)\in \mathbb{L}$ dedicated to UL (DL) traffic.
\subsubsection{Auxiliary variables: link and subchannel activity indicators $J_{ij}, J_{f}$} The binary variables $J_{ij} \in \{0,1\}$ indicate the usage of the link $(i,j) \in \mathbb{L}$, where the binary variable $J_{f} \in \{0,1\}$ indicates the activeness of a frequency subchannel. A wireless link (subchannel) is active iff the transmit power associated with any of the associated subchannels (links) is non-zero. 
%%\subsubsection{Auxiliary variables; achievable information rate $C_{ij,f}$} 
\begin{table}[!t]  
	\renewcommand{\arraystretch}{1.1}
  \caption{Set of decision variables}\label{FD_Cellular_tab_dec_vars}
  \centering
  \begin{tabular}[t]{||c|l||}
  %\firsthline
	\hline
   Set & Description \\
   \hline
   $\mathbb{P}$        &  Set of the variables $X_{ij,f},\; \forall (i,j)\in \mathbb{L}, \; f \in \mathbb{F}$ \\
	 $\mathbb{C}_U (\mathbb{C}_D)$        &   Set of the variables $C_{\text{ul},ij} (C_{\text{dl},ij}), \forall (i,j)\in \mathbb{L} $\\
   $\mathbb{J}_L (\mathbb{J}_F)$        &  Set of the all link (subchannel) activity indicator binary variables \\
	\hline
  \end{tabular}
\end{table} 

\subsection{Network cost model} \label{FD_Cellular_planning_costmodel}
The main required expenditures for the successful function of the wireless backhaul network can be expressed into the following three parts: 
\subsubsection{Power consumption} The cost of power can be formulated as $W_p \sum_{(i,j) \in \mathbb{L}} \sum_{f \in \mathbb{F}} X_{ij,f}$, where $W_p$ is the price of power. Note that the value of $W_p$ may be chosen with the direct consideration of the energy market price or with an extra emphasis on energy saving in order to reduce the consequent $\text{CO}_2$ emissions, which is a rising criteria for the design of wireless networks~\cite{7446253, 5978416}. 
\subsubsection{Wireless links} The establishment of the wireless links is considered as one of the main capital expenditure of a wireless backhaul network, including the establishment of directive antennas suitable for backhaul links, as well as the corresponding transmit and receive chains. This can be expressed as $W_l \sum_{(i,j) \in \mathbb{L}}  J_{ij}$, where $W_l$ is the overall cost of establishing a wireless link.    
\subsubsection{Spectrum usage} With almost all of the sub-$6$ GHz spectrum already utilized, obtaining a dedicated spectrum license is considered as a major cost of a wireless network, motivating research regarding efficient spectrum utilization as well as the usage of higher frequency bands~\cite{hur2013millimeter,li2015optimization}. The overall cost of spectrum can be expressed as $W_f \sum_{f \in \mathbb{F}\setminus \mathbb{F}_a } J_f $ where $W_f$ represent the cost of spectrum for each frequency subchannel dedicated to backhaul. \par
Consequently, the collective network cost is expresses in relation to the decision variables as
\begin{align} \label{FD_Cellular_Aux_Consts}
 & \mathcal{V}\left( \mathbb{P}, \mathbb{J}_L, \mathbb{J}_F\right) = \nonumber \\ & W_p \sum_{(i,j) \in \mathbb{L}} \sum_{f \in \mathbb{F}} X_{ij,f} + W_l \sum_{(i,j) \in \mathbb{L}}  J_{ij} + W_f \sum_{f \in \mathbb{F}\setminus \mathbb{F}_a } J_f. 
\end{align}

\subsection{Constraints}
Formulation of the system constraints are essential in order to ensure a feasible solution; considering the operational limits, e.g., maximum transmit power, as well as the service requirements, e.g., successful transportation of the required UL and DL traffic under an acceptable delay range. 
\subsubsection{Auxiliary variables} 
The value of the defined auxiliary variables are inferred from the main decision variables $X_{ij,f}$, via the imposition of the following constraints:
\begin{flalign} \label{FD_Cellular_Aux_Consts}
&\text{C0:}\;\;\;\;\; X_{ij,f} \geq 0, \forall (i,j) \in \mathbb{L},\; f \in \mathbb{F},   \nonumber  \\
&\text{C1:}\;\;\;\;\; \sum_{f \in \mathbb{F}} X_{ij,f} \; \leq J_{ij} \tilde{P}_{\text{max}},\;  J_{ij} \in \{0,1\}, \; \forall (i,j) \in \mathbb{L},   \nonumber  \\
&\text{C2:}\;\;\; \sum_{(i,j) \in \mathbb{L}} X_{ij,f}  \leq J_{f} \tilde{P}_{\text{max}} ,\;  J_{f} \in \{0,1\}, \; \forall f \in \mathbb{F}, \nonumber 
\end{flalign}
where C0 enforces the domain of the power values, and $\tilde{P}_{\text{max}}$ is any arbitrary upper bound on the total network power consumption, e.g., $\tilde{P}_{\text{max}} = \sum_{i \in \mathbb{N}} P_{\text{max},i}$. The value of $J_{ij}$ ($J_{f}$) is $1$ if any of the corresponding power values are non-zero. Moreover, they are forced to $0$, if the corresponding link (frequency subchannel) is inactive, to reduce the cost function.   
\subsubsection{Transmit power constraints} The per-link power constraint as well as the constraint on the power budget at each node are respectively formulated as
\begin{align} % \label{FD_Cellular_Pow_Consts}
&\text{C3:}\;\;\; \sum_{f \in \mathbb{F}} X_{ij,f} \leq  P_{\text{max},ij},\;\;  \forall (i,j) \in \mathbb{L},  \nonumber  \\
&\text{C4:}\;\;\; \sum_{(i,{x}) \in \mathbb{L}} \sum_{f \in \mathbb{F}} X_{i{x},f} \leq P_{\text{max},i} ,\;\; \forall i \in \mathbb{N}, \nonumber 
\end{align}
see Subsection~\ref{FD_Cellular_model_net_info}.

\subsubsection{Average delay constraint}  
Related to many applications, latency is considered as an important feature of communication quality in the context of next generation communication systems~\cite{7374694}. The delay associated with the transfer of information from (to) the core network to (from) the backhaul nodes is dominated by the processing delay, see Subsection~\ref{FD_Cellular_model_net_info}. In this regard, the average network latency is formulated in relation to the per-node delay as  
\begin{align} % \label{FD_Cellular_delay_Consts}
& \text{C5:}\;\;\; \frac{ \sum_{i \in \mathbb{N}} d_i \sum_{(i,x) \in \mathbb{L}} C_{\mathcal{X},ix}  }{ \sum_{i \in \mathbb{N}} R_{\mathcal{X},i} } \leq \bar{d}_{\mathcal{X}}, \;\;  \mathcal{X} \in \{ul, dl\}, \nonumber    
%\text{C4:}\;\;\; \sum_{X \in \mathbb{N}} \sum_{(i,X) \in \mathbb{L}} \sum_{f \in \mathbb{F}} X_{iX}^{[f]}  \leq P_{\text{max}}^{[i]}  ,\;\; \forall i \in \mathbb{N},
\end{align}
where $\bar{d}_{ul}$ ($\bar{d}_{dl}$) is the tolerable average UL (DL) delay. 

\subsubsection{Interference threshold on access network}
The coexistence of the backhaul and access network is conditioned on complying with the tolerable collective interference threshold $I_{\text{th},l,f}$, see Subsection~\ref{FD_Cellular_model_net_info}. This is expressed as  
\begin{flalign} % \label{FD_Cellular_access_intrf_threshold_Consts}
&\text{C6:}\;\;\; \sum_{(i,j)\in \mathbb{L}}  \Omega_{ij,l,f} X_{ij,f} \leq I_{\text{th},l,f},\;\; \forall f \in \mathbb{F}_a, \; l \in \mathbb{N}. \nonumber
\end{flalign}

\subsubsection{Link information flow constraint}
The physical information capacity on each link, including the wired and wireless connections, need to be sufficient for the dedicated UL and DL information rates, i.e., $C_{\text{ul},ij}, C_{\text{dl},ij}$. This is expressed as  
\begin{flalign} % \label{FD_Cellular_link_capacity_Consts}
\text{C7:}\;\;\; C_{\text{ul},ij} + C_{\text{dl},ij} \leq C_{0,ij} + \sum_{f \in \mathbb{F}} C_{ij,f}, \;\; \forall (i,j) \in \mathbb{L},  \nonumber 
\end{flalign}
where $C_{ij,f}$ is related to the other variables from (\ref{FD_Cellular_model_capacity_per_link}), and $C_{0,ij}$ indicates the available link capacity via wired technologies. 

\subsubsection{Network information flow constraint}
The preservation of the network information flow, separately at the root nodes and non root nodes are formulated as 
\begin{flalign} % \label{FD_Cellular_inf_flow_nonroot_Consts}
&\text{C8:}\;\;\;\;\; R_{\text{ul},i} \;= \sum_{ (i,x) \in \mathbb{L}} C_{\text{ul},ix}  - \sum_{ (y,i) \in \mathbb{L}} C_{\text{ul},yi}  , \;\; \forall i \in \mathbb{M},  \nonumber  \\
&\text{C9:}\;\;\;\;\; R_{\text{dl},i} \;= \sum_{ (x,i) \in \mathbb{L}} C_{\text{dl},xi}  - \sum_{ (i,y) \in \mathbb{L}} C_{\text{dl},iy}, \;\; \forall i \in \mathbb{M},  \nonumber  \\
&\text{C10:}\;\;\; \sum_{i\in \mathbb{M}} R_{\text{ul},i} = \sum_{i\in \mathbb{R}} \sum_{  (j,i) \in \mathbb{L}} C_{\text{ul},ji}, \nonumber  \\
&\text{C11:}\;\;\; \sum_{i\in \mathbb{M}} R_{\text{dl},i} = \sum_{i\in \mathbb{R}} \sum_{  (i,j) \in \mathbb{L}} C_{\text{dl},ij},  \nonumber  
\end{flalign}
where C8-9 indicate the flow conservation at each node, and C10-11 represent the conservation of the information over the network. 

%%%%%%%%%%%%%%%%%%%%%%%%%%%%%%%%%%%%%%%%%%%%%%%%%%%%%%%%%%%%%%%%%%%%%%%%%%%%%%%
%%%%%%%%%%%%%%%%%%%%%%%%%%%%%%%%%%%%%%%%%%%%%%%%%%%%%%%%%%%%%%%%%%%%%%%%%%%%%%%
%%%%%%%%%%%%%%%%%%%%%%%%%%%%%%%%%%%%%%%%%%%%%%%%%%%%%%%%%%%%%%%%%%%%%%%%%%%%%%%
%%%%%%%%%%%%%%%%%%%%%%%%%%%%%%%%%%%%%%%%%%%%%%%%%%%%%%%%%%%%%%%%%%%%%%%%%%%%%%%
%%%%%%%%%%%%%%%%%%%%%%%%%%%%%%%%%%%%%%%%%%%%%%%%%%%%%%%%%%%%%%%%%%%%%%%%%%%%%%%

\subsection{Network planning: an MILP model} \label{FD_Cellular_sec_MILP_opt}
In this part we provide an MILP framework, addressing a minimum cost wireless backhaul network design. The corresponding optimization problem is formulated as 
\begin{subequations}  \label{FD_Cellular_Opt_MILP_min_cost_0}
\begin{align}
\underset{\mathbb{P}, \mathbb{C}_{U}, \mathbb{C}_{D}, \mathbb{J}_L , \mathbb{J}_F }{\text{minimize}} \;\; \hspace{-2mm}  & \mathcal{V}\left( \mathbb{P}, \mathbb{J}_L, \mathbb{J}_F\right)  \;\;\;  {\text{s.t.}} \;\; \text{C0-C11},
\end{align}
\end{subequations}
where the sets $\mathbb{P}, \mathbb{C}_{U}, \mathbb{C}_{D}, \mathbb{J}_L , \mathbb{J}_F $ represent the decision variables, see Table~\ref{FD_Cellular_tab_dec_vars}. It is observed that the cost function, as well as the constraints C0-C6 and C8-C11 comply with the intended mixed linear structure. However, the above problem is not an MILP due to the non-linear constraint C7. In order to observe this, we recall from (\ref{FD_Cellular_model_capacity_per_link}) that the achievable information rate at each link is related to the power of the desired and interfering links as    
\begin{align} \label{FD_Cellular_link_capacity_formulation_2}
C_{ij,f} = f_1 (\mathbb{P}) - f_2 (\mathbb{P}),    
\end{align} 
where 
\begin{align}
f_1 (\mathbb{P}) &:= B \text{log}\left( \sum_{(l,k) \in \left\{ \mathbb{L}\right\} } \Gamma_{ij,lk,f} X_{lk,f} + W_{ij,f} \right), \\
f_2 (\mathbb{P}) &:= B \text{log}\left( \sum_{(l,k) \in \left\{ \mathbb{L} \setminus (i,j) \right\} } \Gamma_{ij,lk,f} X_{lk,f} + W_{ij,f} \right),
\end{align} %%%%%the achievable rate variables, which relate to the power variables $X_{i{x},f}$  with the exception of the C7. However, the constraint C7, An optimization problem for a minimum-cost planning for the defined system can be expressed as 
%%%%%
%%%%%It is observed that with the exception of the physical link capacity expression (\ref{FD_Cellular_model_capacity_per_link}), the defined cost model in Subsection~\ref{FD_Cellular_planning_costmodel} as well as the constraints C1-C11 can be well modeled as an MILP. It is worth mentioning that a realistic consideration of wireless link capacity, in connection to the desired and interference paths, is essential for a meaningful implementation of an spectrum/interference management scheme. The role of the variables $C_{ij,f}$ is hence expressed as $C_{ij,f} = f_1 (\mathbb{P}) - f_2 (\mathbb{P})$, see (\ref{FD_Cellular_planning_link_capacity_Consts_2}), %\begin{align} \label{FD_Cellular_planning_link_capacity_Consts}
%%%%%%C_{ij,f} = f_1 (\mathbb{P}) - 
%%%%%%\end{align}  
%%%%%%\begin{align} \label{FD_Cellular_planning_link_capacity_Consts}
%%%%%%C_{ij,f} = & B \log_2 \left( \Lambda_{ij,f} X_{ij,f} + \hspace{-3mm} \sum_{(l,k) \in \left\{ \mathbb{L} \setminus (i,j) \right\} } \hspace{-3mm} \Gamma_{ij,lk,f} X_{lk,f} + W_{ij,f} \right)  \nonumber \\ & - B \log_2 \left(\sum_{(l,k) \in \left\{ \mathbb{L} \setminus (i,j) \right\} }\hspace{-5mm}  \Gamma_{ij,lk,f} X_{lk,f}  + W_{ij,f} \right),
%%%%%%\end{align}
which holds a difference of concave functions over the decision variables $\mathbb{P}$. Unfortunately, this formulation is challenging for the standard numerical solvers due to \emph{i)} the non-convexity of the resulting feasible set in C7, and \emph{ii)} the logarithmic concave-convex expressions. In order to comply with the intended MILP framework, we undertake three steps. Firstly, we introduce the achievable link capacity $C_{ij,f}$ as an auxiliary variable in the optimization and directly impose (\ref{FD_Cellular_link_capacity_formulation_2}) as a constraint, thereby linierizing the constraint set C0-C11. Secondly, the equality in (\ref{FD_Cellular_link_capacity_formulation_2}) is relaxed as an inequality constraint, however, it is easily verified that the constraint will be tight for an optimal system design, resulting in zero relaxation gap\footnote{The proof is obtained via contradiction; if for an optimal design of network parameters the constraint is not tight for the link $(i,j)$ and at the subchannel $f$, then the transmit power value $X_{ij,f}$ can reduced until the constraint is tight. This will reduce the objective (cost), while does not violate any of the design constraints.}. And third, we apply linear conservative approximations on the non-linear terms $f_1$ and $f_2$. In this respect, the logarithmic term $f_1$ is approximated as a piecewise linear function, i.e., approximating the concave expression as a maximum of multiple affine functions, denoted as $\mathcal{L}_p \left( \mathbb{P} ; \mathbb{P}_0 \right)$, such that 
\begin{align} \label{FD_Cellular_f_1_upper_approx}
\mathcal{L}_p \left( \mathbb{P} ; \mathbb{P}_0 \right) \leq f_1 (\mathbb{P}), \;\; \mathcal{L}_p \left( \mathbb{P}_0 ; \mathbb{P}_0 \right) = f_1 (\mathbb{P}_0). 
\end{align}
Moreover, the logarithmic term $f_2$ is approximated as an over-estimating affine, denoted as $\mathcal{L} \left( \mathbb{P} ; \mathbb{P}_0 \right)$, such that 
\begin{align} \label{FD_Cellular_f_2_upper_approx}
\mathcal{L} \left( \mathbb{P} ; \mathbb{P}_0 \right) \geq f_2 (\mathbb{P}), \;\; \mathcal{L} \left( \mathbb{P}_0 ; \mathbb{P}_0 \right) = f_2 (\mathbb{P}_0), 
\end{align}
which is directly obtained via the first-order Taylor's expansion at the point $\mathbb{P}_0$\footnote{A tight affine approximation of a convex (concave) function obtained via Taylor's approximation, is also a global lower (upper) bound \cite{BV:04}.}. Note that the collective approximation generates a tight and piecewise affine function, which bounds the original nonlinear function from above, see $(b)$. The wireless link capacity constraint can be hence satisfied by imposing
%approximation the difference to the    
%Note that (\ref{FD_Cellular_planning_link_capacity_Consts}) intends to ensure the satisfaction of the subchannel capacity value $C_{ij,f}$, given the power of the wireless links $\mathbb{P}$. In order to comply with the intended MILP structure, we apply a conservative approximation of the second logarithmic term, obtained by applying the Taylor's approximation on the convex part. Moreover, we approximate the concave part by replacing the logarithm with a conservative piecewise linear function, i.e., approximating the concave expression as a maximum of multiple affine expressions. The inequality (\ref{FD_Cellular_}) can be hence approximated as 
\begin{align} \label{FD_Cellular_link_capacity_Consts_3}
C_{ij,f} \leq  \mathcal{L}_p \left( \mathbb{P}_0 , \mathbb{P} \right) - \mathcal{L}\left( \mathbb{P}_0 , \mathbb{P} \right), 
\end{align} 
where $\mathbb{P}_0$ is the point of approximation. Note that the satisfaction of (\ref{FD_Cellular_link_capacity_Consts_3}) consequently ensures that the wireless link realizes the capacity value $C_{ij,f}$, due to the proposed conservative approximation. However, this may result in an inefficient solution, due to the approximation gap\footnote{A large deviation of approximated piecewise linear function $\mathcal{L}_p - \mathcal{L}$ with the original nonlinear expression results in the under-utilization of the wireless link capacity, and consequently additional costs.}. In this regard, an iterative update is applied, where the obtained variable set $\mathbb{P}$ at each iteration is set as the approximation point, i.e., $\mathbb{P}_0$, for the next design iterations. The minimum-cost network design problem can be hence formulated as
\begin{subequations}  \label{FD_Cellular_Opt_MILP_min_cost_final}
{{\begin{align}
\underset{\mathbb{V}^{[m]}}{\text{minimize}} \;\;   &  \mathcal{V}\left( \mathbb{P}^{[m]}, \mathbb{J}_L^{[m]}, \mathbb{J}_F^{[m]} \right)  \label{FD_Cellular_eq:global_opt_problem_MWMSE_a} \\
{\text{s.t.}} \;\;\;\;\;\;\;\;\;\; & C_{ij,f} \leq  \mathcal{L}_p \left(  \mathbb{P}^{[m]} ; \mathbb{P}^{[m-1]} \right) - \mathcal{L}\left( \mathbb{P}^{[m]} ; \mathbb{P}^{[m-1]} \right) , \nonumber \\ &  \;\;\;\;\;\;\;\;\;\;\;\;\;\;\;\;\;\;\;\;\;\;\;\;\;\;\;\;\;\;\; \forall (i,j) \in \mathbb{L},\;\; \forall f \in \mathbb{F}, \\
 & \text{C0-C11}, \label{FD_Cellular_eq:global_opt_problem_MWMSE_b}
\end{align} }} 
\end{subequations}where 
\begin{align} 
\mathbb{V}^{[m]}:= \left\{\mathbb{P}^{[m]}, \mathbb{C}_{U}^{[m]}, \mathbb{C}_{D}^{[m]}, \tilde{\mathbb{C}}^{[m]}, \mathbb{J}_L^{[m]} , \mathbb{J}_F^{[m]} \right\},
\end{align} 
the set $\tilde{\mathbb{C}}$ represents the sets of the variables $C_{ij,f}$, and $m$ denotes the iteration index, see Algorithm~\ref{FD_Cellular_alg_1}. The algorithm stops as a stable set of decision variables is obtained, or a maximum number of iterations is expired. 
\subsubsection{Numerical implementation} \label{FD_Cellular_alg_numericalImpl}
As intended, the obtained optimization framework (\ref{FD_Cellular_Opt_MILP_min_cost_final}) is an MILP in each iteration. Note that due to the combinatorial structure, such problems are not convex, and hence the popular interior point methods may not be applied \cite{BV:04}. However, efficient variations of branch and bound methods \cite{bertsekas1999nonlinear} have been recently developed and implemented in the framework of the standard numerical solvers, e.g., CPLEX, Gurobi, resulting in efficient numerical solutions for the MILP problems with large dimensions.
\subsubsection{Algorithm initialization} \label{FD_Cellular_alg_init}
The algorithm starts by activating all feasible physical links, frequency sub channels, as well as the maximum power consumptions at all wireless links, respecting the constraints C0-2. Note that this initialization choice corresponds to a strong link and network capacity. However, it corresponds with the maximum utilization of the network resources, resulting in the maximum cost. 
%As defined in Algorithm~\ref{FD_Cellular_alg_1}, branch-and-bound solver iterations as well as the approximation updates are consequently applied to reduce the cost. 
\subsubsection{Convergence} \label{FD_Cellular_alg_conv}
The convergence behavior of the algorithm is of interest, due to the proposed iterative cost reduction. Via to the application of the branch and bound update by the numerical solver, the solution experiences a monotonic enhancement in each iteration, within a fixed tolerance region. Note that such monotonic reduction of the cost function holds at the internal solver iterations, as well as the external iterations by updating the approximations, see Algorithm~\ref{FD_Cellular_alg_1}, step~$4$. This results in a necessary algorithm convergence, due to the fact that the problem objective is bounded from below. Further analysis regarding the algorithm convergence behavior and computational complexity is conducted via numerical simulations in Section~\ref{sec_simulations}. 
%However, the global optimality of the obtained solution may not be guaranteed due to \emph{i)} the non-convex nature of the combinatorial problems, and \emph{ii)} possibility of converging into a local optimum. Further analysis regarding the algorithm convergence behavior and computational complexity is conducted via numerical simulations in Section~\ref{FD_Cellular_}. 
%%\subsubsection{Computational complexity}
%%
 %%However, the global optimality of the obtained may not be guaranteed due to the i) nonö-convex nature of the MILP structure, where a global optimality may only be guaranteed via the utilization of an exhaustive search over the feasible variable space, and ii) the itarative update of the optimization variables, and the possibility of a local optimum solution. The convergence behavior of the XXXX algorithm is studied via numerical simulations in XXX.  
%%Further analysis regarding the algorithm convergence behavior and computational complexity is conducted via numerical simulations in Section~\ref{FD_Cellular_}. as well as the 
%%
%
\begin{algorithm}[H] 
 \small{	\begin{algorithmic}[1] 
  %\SetAlgoLined
\State{$m \gets  {0} ; \;\; \mathbb{P}^{[0]} \leftarrow  \text{Subsection~\ref{FD_Cellular_alg_init}};$}  \Comment{initialization}
\Repeat 
\State{$m \leftarrow  m + 1;$}
\State{$\mathbb{P}^{[m]},\mathbb{C}_{U}^{[m]}, \mathbb{C}_{D}^{[m]}, \tilde{\mathbb{C}}^{[m]}, \mathbb{J}_L^{[m]},\mathbb{J}_F^{[m]} \leftarrow  \text{solve~MILP~(\ref{FD_Cellular_Opt_MILP_min_cost_final})} ;$}
\Until{$\text{a stable point, or a maximum number of $m$ reached}$}
\State{\Return$\left\{\mathbb{P}^{[m]}, \mathbb{C}_{UL}^{[m]}, \mathbb{C}_{DL}^{[m]}, \mathbb{J}_L^{[m]} , \mathbb{J}_F^{[m]} \right\}$}
  \end{algorithmic} } 
 \caption{\small{MILP-based minimum cost network planning.} } \label{FD_Cellular_alg_1}
\end{algorithm}  

\subsection{Network re-tuning: an SIA model}
The joint consideration of the power allocation on wireless links as well as the frequency and topology planning is expected to be gainful as the aforementioned factors jointly impact the network interference pattern and information flow. In this regard, a full scale design of the wireless backhaul network is proposed in the previous part, assuming the availability of the accurate network information and integrating the FD capability. In this part, we propose a methodology to re-tune the network operation by adjusting the transmit power at the wireless links, assuming that the links, as well as the operating frequencies are already established. In particular, this approach enables the network to adapt to slight changes, e.g., change in the channel situations due to weather and temperature fluctuations, a slight increase or decrease in the required information rate, and the occasional degradation of the SIC quality. Moreover, contrary to usual planning strategies which focus on the worst-case network requirements, the provided flexibility results in a higher efficiency and reduced cost, e.g., by reducing the transmit power at a wireless link when the data traffic is low. And finally, the reduced setup complexity, as a result of the fixed network topology, is constructively used to obtain a computationally more efficient design. 

%Since the wireless links as well as the operating frequency bands are already setup, prior to the re-tuning process, 
Given an active set of links and frequency subchannels, the problem of adjusting the transmit powers at the wireless links is formulated as
%\begin{subequations}  \label{FD_Cellular_Opt_SIC_re_tunning_0}
%{\small{\begin{align}
%\underset{\mathbb{P}, \mathbb{C}_{UL}, \mathbb{C}_{DL} }{ \text{minimize}} \;\; \hspace{-0mm}  & \sum_{(i,j) \in \mathbb{L}} \sum_{f \in \mathbb{F}} X_{ij,f}  \label{FD_Cellular_eq:global_opt_problem_MWMSE_2a} \\
%{\text{s.t.}} \;\;\;\;\;\;\;\; & C_{ij}^{[f]} \leq  f_1 \left(  \mathbb{P}^{[m]} \right) - \mathcal{L}\left( \mathbb{P}^{[m]} ; \mathbb{P}^{[m-1]} \right) , \\
 %& \text{C3-C12}, \label{FD_Cellular_eq:global_opt_problem_MWMSE_2b}
%\end{align} }}
%\end{subequations}  
%\begin{subequations} 
\begin{align} \label{FD_Cellular_Opt_SIC_re_tunning}
\underset{ \mathbb{P}, \mathbb{C}_{UL}, \mathbb{C}_{DL} }{\text{minimize}} \;\; \hspace{-0mm}  \sum_{(i,j) \in \mathbb{L}} \sum_{f \in \mathbb{F}} X_{ij,f}, \;\; \text{C0,~~C3-C11}. 
\end{align} 
%\end{subequations}
It is observed that due to the elimination of the binary variable sets $\mathbb{J}_L,\mathbb{J}_F$, the problem is simplified to a non-linear program over a continuous domain, hence, the utilization of MILP-based solvers is not necessary. However, the problem (\ref{FD_Cellular_Opt_SIC_re_tunning}) it is not a convex optimization problem due to the constraint C7. In this regard, we introduce an {SIA} framework~\cite{marks1978technical}, where the non-convex feasible region constructed by C7 is inner-approximated in each iteration as a convex set. In this regard, we follow the same procedure as implemented for (\ref{FD_Cellular_Opt_MILP_min_cost_final}), i.e., introduction and relaxation of (\ref{FD_Cellular_link_capacity_formulation_2}) as an inequality constraint, and upper-approximation of $f_2$ via (\ref{FD_Cellular_f_2_upper_approx}). However, the term $f_1$ can be directly applied since it does not violate the convexity. The approximated optimization problem is hence formulated as  
\begin{subequations}  \label{FD_Cellular_Opt_SIC_re_tunning_final}
\begin{align}
\underset{\mathbb{P}^{[m]}, \tilde{\mathbb{C}}^{[m]}, \mathbb{C}_{U}^{[m]}, \mathbb{C}_{D}^{[m]} }{ \text{minimize}} \;\;   &  \sum_{(i,j) \in \mathbb{L}} \sum_{f \in \mathbb{F}} X_{ij,f}   \\
{\text{s.t.}} \;\;\;\;\;\;\;\;\;\; & C_{ij,f} \leq  f_1 \left(  \mathbb{P}^{[m]} \right) - \mathcal{L}\left( \mathbb{P}^{[m]} ; \mathbb{P}^{[m-1]} \right) , \nonumber \\ & \;\;\;\;\;\;\;\;\;\;\;\;\;\;\;\;\;\;\;\;\;\;\;\;\;\;\;\;\;\; \forall (i,j) \in \mathbb{L},\;\; \forall f \in \mathbb{F}, \\
 & \text{C0,~~C3-C11},
\end{align} 
\end{subequations}\hspace{-0mm}
where $m$ denotes algorithm iteration. The iterations of (\ref{FD_Cellular_Opt_SIC_re_tunning_final}) are continued until convergence, or a maximum number of algorithm iterations is expired, see Algorithm~\ref{FD_Cellular_alg_2} for a detailed procedure.

\begin{algorithm}[H] 
 \small{	\begin{algorithmic}[1] 
  %\SetAlgoLined
\State{$m \gets  {0} ; \;\; \mathbb{P}^{[0]} \leftarrow  \text{Subsection~\ref{FD_Cellular_alg_init_retunning}};$}  \Comment{initialization}
\Repeat 
\State{$m \leftarrow  m + 1;$}
\State{$\mathbb{P}^{[m]} \leftarrow  \text{solve~(\ref{FD_Cellular_Opt_SIC_re_tunning})} ;$}
\Until{$\text{a stable point, or a maximum number of $m$ reached}$}
\State{\Return$\left\{\mathbb{P}^{[m]}, \mathbb{C}_{U}^{[m]}, \mathbb{C}_{D}^{[m]} \right\}$}
  \end{algorithmic} } 
 \caption{\small{SIC-based network re-tuning.} } \label{FD_Cellular_alg_2}
\end{algorithm}     

\subsubsection{Algorithm initialization} \label{FD_Cellular_alg_init_retunning}
Since the re-tuning algorithm is intended to operate on an existing network, the current state of the network including the transmit power of the wireless links is used as the initial point. In case such initial point may not be obtained, a similar initialization as introduced in Subsection~\ref{FD_Cellular_alg_init} can be used. 
\subsubsection{Convergence and numerical implementation} \label{FD_Cellular_alg_numericalImpl_retunning}
Due to the convexity of the approximated problem (\ref{FD_Cellular_Opt_SIC_re_tunning_final}) in each iteration, the optimum solution can be obtained via standard numerical solvers for convex problems, e.g., SeDuMi~\cite{sturm1999using}, SDPT3~\cite{tohsdpt3}, employing efficient interior point methods. Moreover, it is easily verified that the proposed linear approximation satisfies the set of properties established in \cite[Theorem~1]{marks1978technical} for smooth problems, i.e., tightness and globally lower bound properties. Together with the fact that the approximated problem at each step is solved to the optimality, it certifies a necessary convergence to a stationary point of the original non-convex problem (\ref{FD_Cellular_Opt_SIC_re_tunning}).

\section {Numerical evaluation}\label{sec_simulations}
\begin{figure}[h] 
    \begin{center}
        \includegraphics[angle=0,width=0.9\columnwidth]{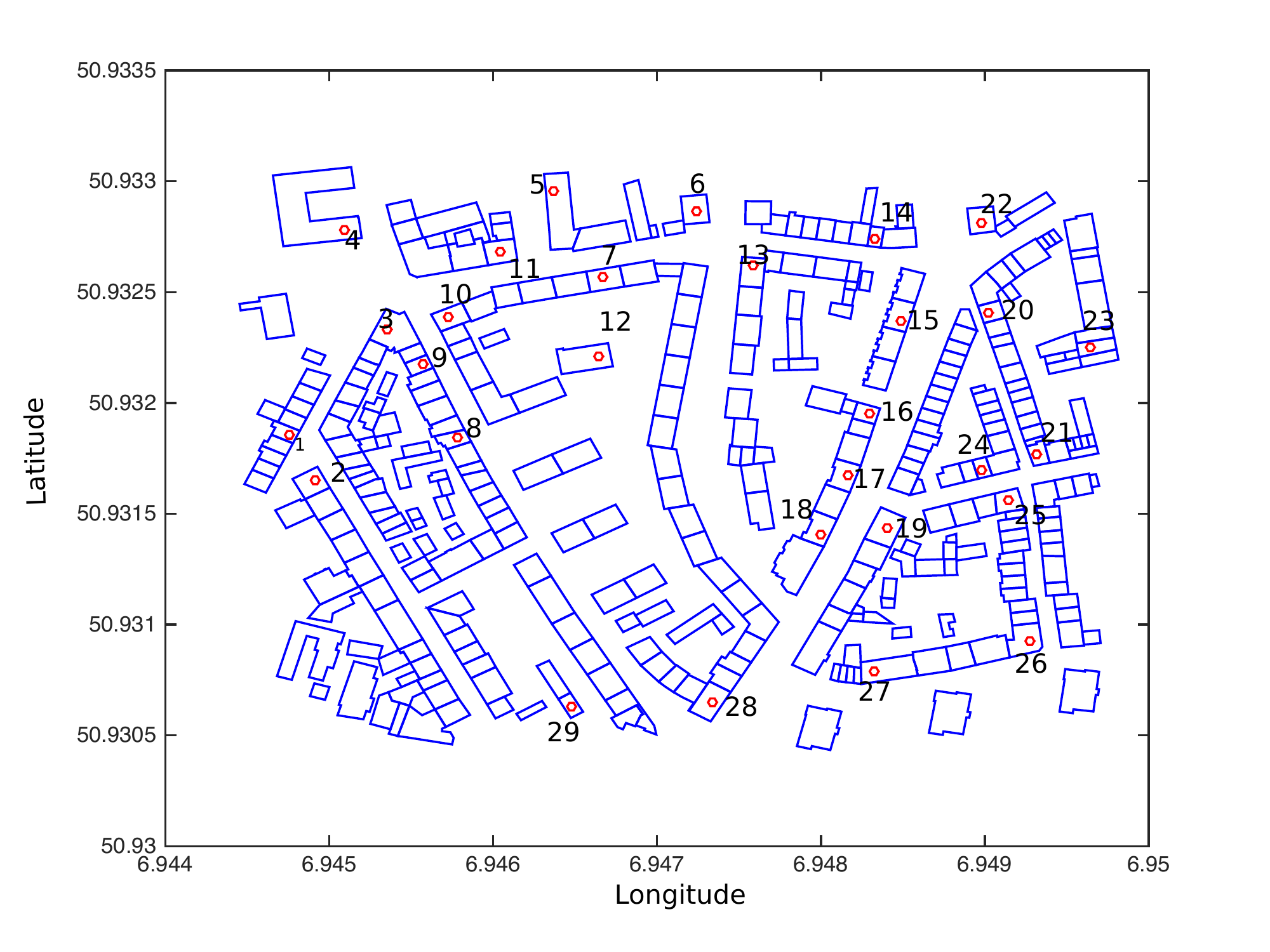} 
				        %\fbox{model_rect6.pdf}
    \caption{The simulated environment from Aachen city map. Red dots represent the location of the backhaul nodes, put on top of the rooftops, whereas blue squares represent the body area of the buildings.} \label{FD_Cellular_fig_aachen_city_backhaul}
    \end{center} 
\end{figure} 

%%\begin{figure}[h] 
    %%\begin{center}
        %%\includegraphics[angle=0,width=0.9\columnwidth]{./Sections/FD_Cellular/Figures/Convergence3} 
				        %%%\fbox{model_rect6.pdf}
    %%\caption{The simulated environment from Aachen city map. Red dots represent the location of the backhaul nodes, put on top of the rooftops, whereas blue squares represent the body area of the buildings.} \label{FD_Cellular_fig_aachen_city_backhaul}
    %%\end{center} 
%%\end{figure} 

%%\begin{table}[ht]
	%%\centering
	%%\caption{Simulation Parameters}	
	%%\label{tabsim}
	%%\begin{tabular}{|c|c|}
	%%\hline
	    %%Carrier frequency  & 5GHz  \\
	    %%\hline
		%%One frequency channel Bandwidth & 20MHz \\
		%%\hline
		%%Antenna Type & Directive \\
		%%\hline
		%%Antenna polarization & Co-polarization \\
		%%\hline
		%%Maximum antenna element gain & 13 dBi \\
		%%\hline
		%%Antenna Height & 20 m \\
		%%\hline
		%%Maximum Transmit Power of one Antenna & 23 dBm \\
		%%\hline
		%%Directivity Angle ($^\circ$) & 15 \\
		%%\hline
		%%Number of time samples & 10 \\
		%%\hline
		%%Antenna element gain pattern & 3D \\
		%%\hline
		%%K-factor & 9\\
		%%\hline
		%%Shadow fading & 4 dB\\
		%%\hline
		%%Mobility km/h & 0-70\\		
	%%\hline	
	%%\end{tabular}
%%\end{table}

In this part, we evaluate the impact of the proposed designs in terms of the total network cost, benefiting from the FD capability at the wireless links, via numerical simulations\footnote{I would like to thank my colleague M.Sc. Jose Angel Leon Calvo for providing the environment-related data, and extracting the simulated channel statistics from the city map of Aachen. Moreover, I would like to thank my former students M.Sc. Praveen Sirvi, and M.Sc. Santosh Prahalada Narasimha for their constructive help in the numerical simulations for Section~\ref{FD_Cellular_sec_MILP_opt}. }. The simulated network is obtained from the city map of Aachen, Germany, see Fig.~\ref{FD_Cellular_fig_aachen_city_backhaul}, employing our developed wave propagation simulator \cite{ScReMa12}, in combination with the urban micro-cell scenario from WINNER II model \cite{CaScXuMa15}. The antenna type is chosen with directivity gain of $13$ dBi and directivity angle of $15$ degrees in accordance with the European Standard \cite{etsi302}. The evaluations of the channel properties are carried by keeping the center frequency of $5$ GHz and with bandwidth of each frequency sub-channel of $20$ MHz. The planning algorithm determines the wireless links to be established, as well as the operating power and frequency bands at each link. Unless otherwise is stated, the following values define the default setup: $W_{ij,f}= -97$~dBm, $B= 20$~MHz, $P_{\text{max},ij}= 30$~dBm, $P_{\text{max},i} = 30$~dBm, $R_{\text{req}} = R_{\text{ul},i}=R_{\text{dl},i} = 100$~Mbits/sec, $|\mathbb{N}|=23$, $|\mathbb{F}|=8$, $|\mathbb{L}|=100$, $C_{0,ij}=0$. The self-interference cancellation is assumed to be perfect, i.e., $\rho_{\text{si}} = \Gamma_{ij,li,f} = 0$. The network cost model is set following \cite{mahloo2014cost, ahmed2013study}, as $W_p=1$, $W_l=20$, $W_f=10$.   

\begin{figure*}[!h]  
\hspace{0.0cm} \subfigure[Convergence]{\includegraphics[angle = 0, width = 0.49\columnwidth]{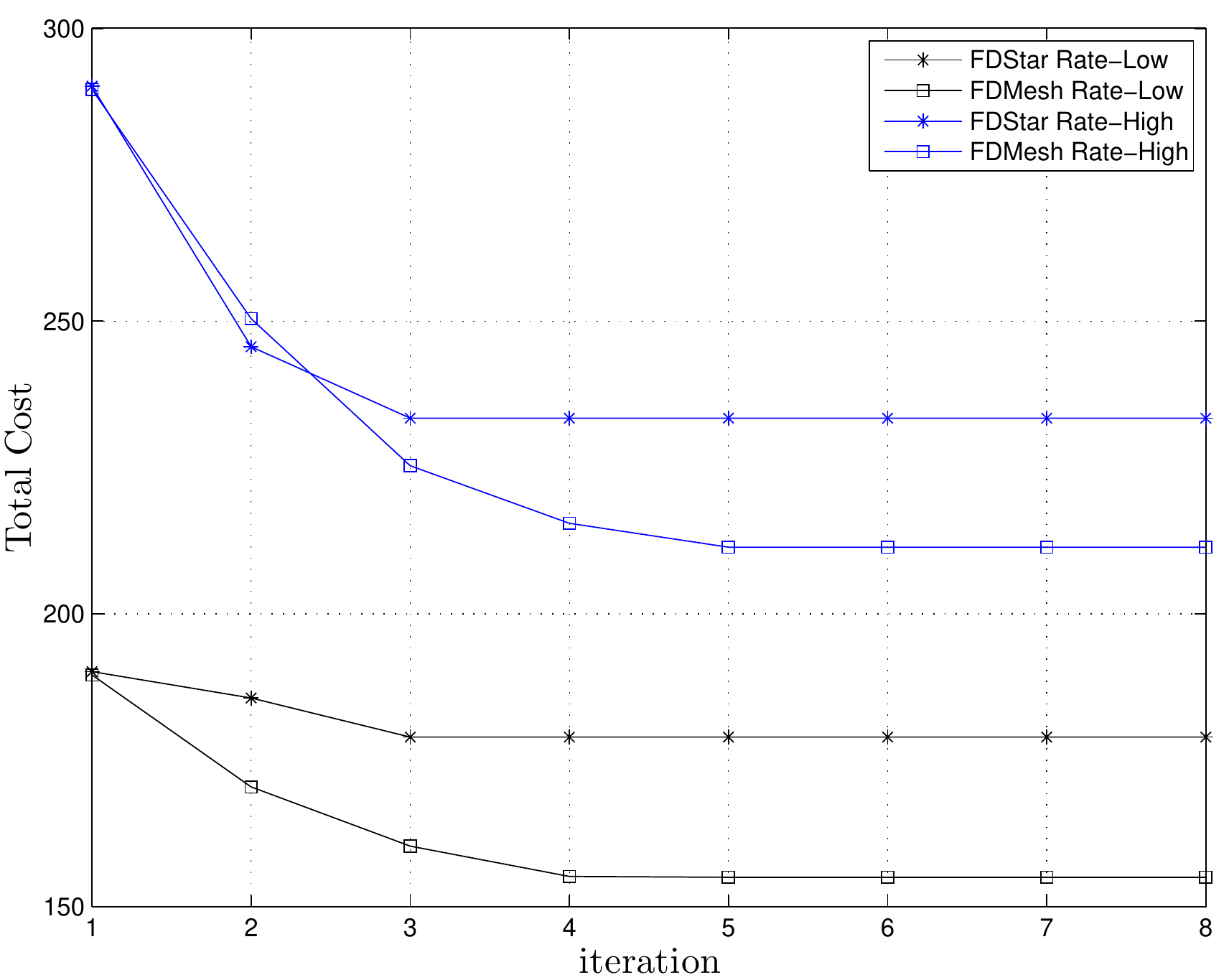}} \label{FD_Cellular_fig_conv1}
\subfigure[Cost vs. SIC level]{\includegraphics[width = 0.49\columnwidth]{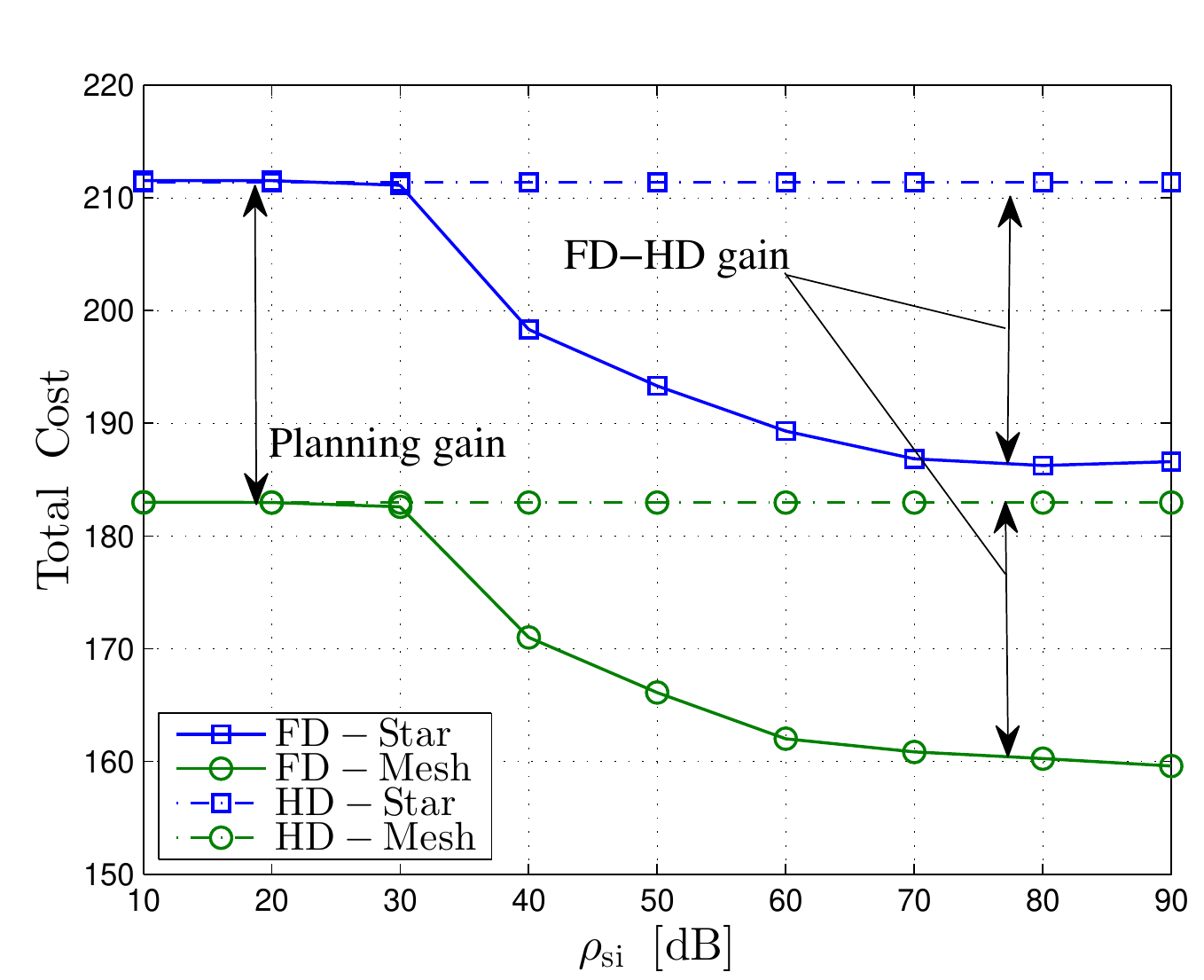}} \label{FD_Cellular_fig_conv2}
\subfigure[Cost vs. $R_{\text{req}}$]{\includegraphics[width = 0.49\columnwidth]{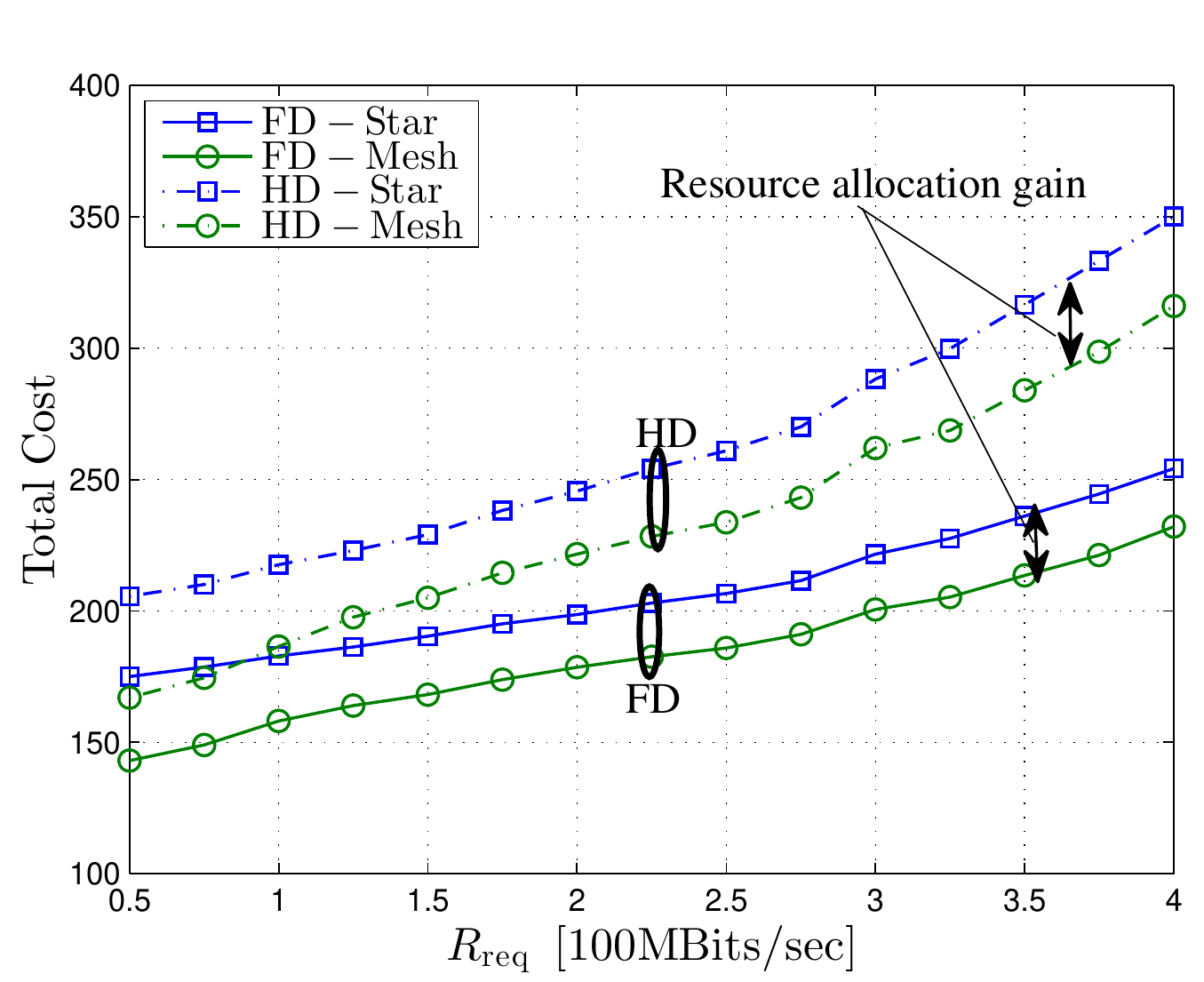}} \label{FD_Cellular_fig_conv3}
\subfigure[Spectrum use vs $R_{\text{req}}$]{\includegraphics[width = 0.49\columnwidth]{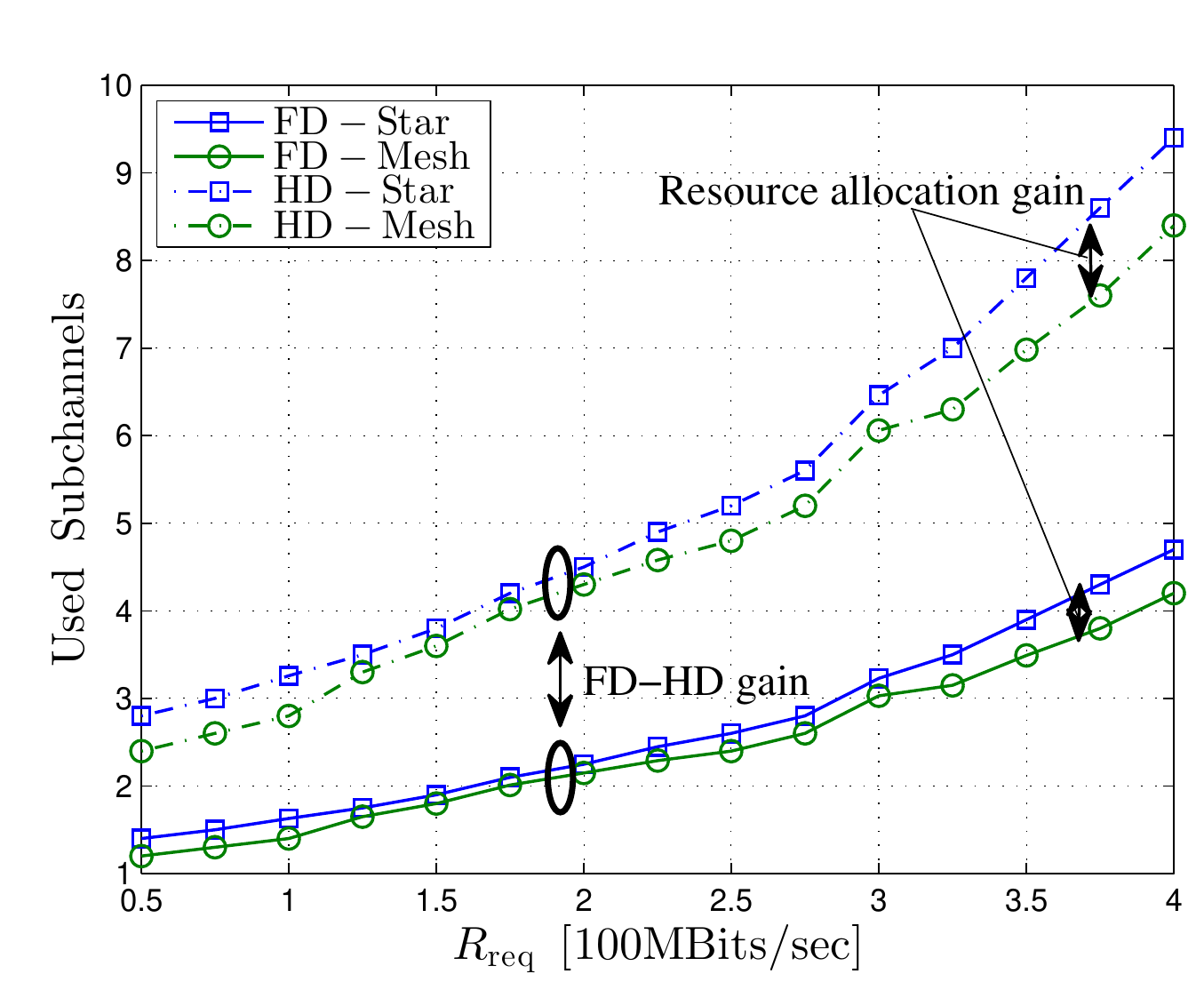}} \label{FD_Cellular_fig_conv2}
\caption{The algorithm convergence behavior (a), and resulting network cost for different values of $R_{\text{req}}$ and $\rho_{\text{si}}$.} \label{FD_Cellular_fig_sim_all}
\end{figure*}

In Fig.~\ref{FD_Cellular_fig_sim_all}~(a), the average convergence behavior of the proposed MILP model is depicted. Due to the proposed iterative improvement, the convergence behavior is interesting as an indication of the algorithm complexity as well as the achievable iterative improvement. The star topology indicates the case where each node is connected to a root node via a direct wireless link, whereas the mesh topology represents the result of the design proposed in Algorithm~\ref{FD_Cellular_alg_1}. Furthermore, the convergence curves indicate the different regimes regarding the backhaul rate requirements, i.e., $R_{\text{req}}=100$, indicated as `Rate-Low', and $R_{\text{req}}=200$, indicated as `Rate-High'. A monotonic decrease in cost is observed, where the convergence is obtained within $3-8$ iterations. 

In Figs.~\ref{FD_Cellular_fig_sim_all}~(b)-(c) the impact of the rate demand is depicted on the collective network cost. It is observed that a higher rate demand results in a higher cost. Moreover, it is observed that the application of FD capability, together with the proposed joint link/frequency planning results in the usage of less frequency subchannels, thereby reducing the total cost by approximately $10-30$\% depending on the traffic load. 

The impact of the SIC quality on the overall network cost is depicted in Fig.~\ref{FD_Cellular_fig_sim_all}~(d). It is observed that the FD gain is reduced as the SIC quality is degraded and reaches the performance of the network with half-duplex links for a poor SIC condition.

\begin{figure*}[!h]  
\hspace{1.5cm} \subfigure[$W_p=0$, $W_l=10$,]{\includegraphics[angle = 0, width = 0.55\columnwidth]{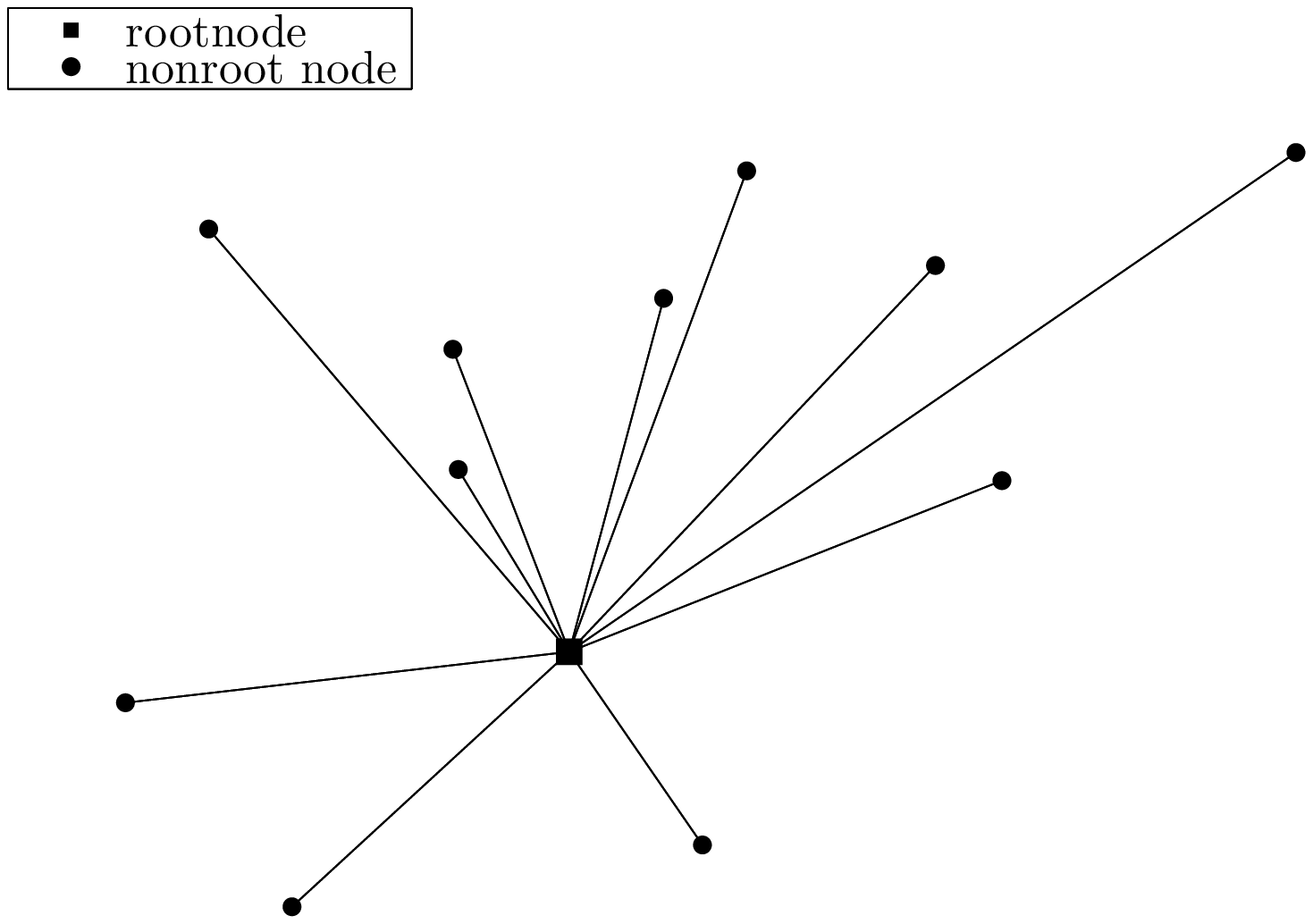}} 
\subfigure[$W_p=10$, $W_l=1$,]{\includegraphics[width = 0.55\columnwidth]{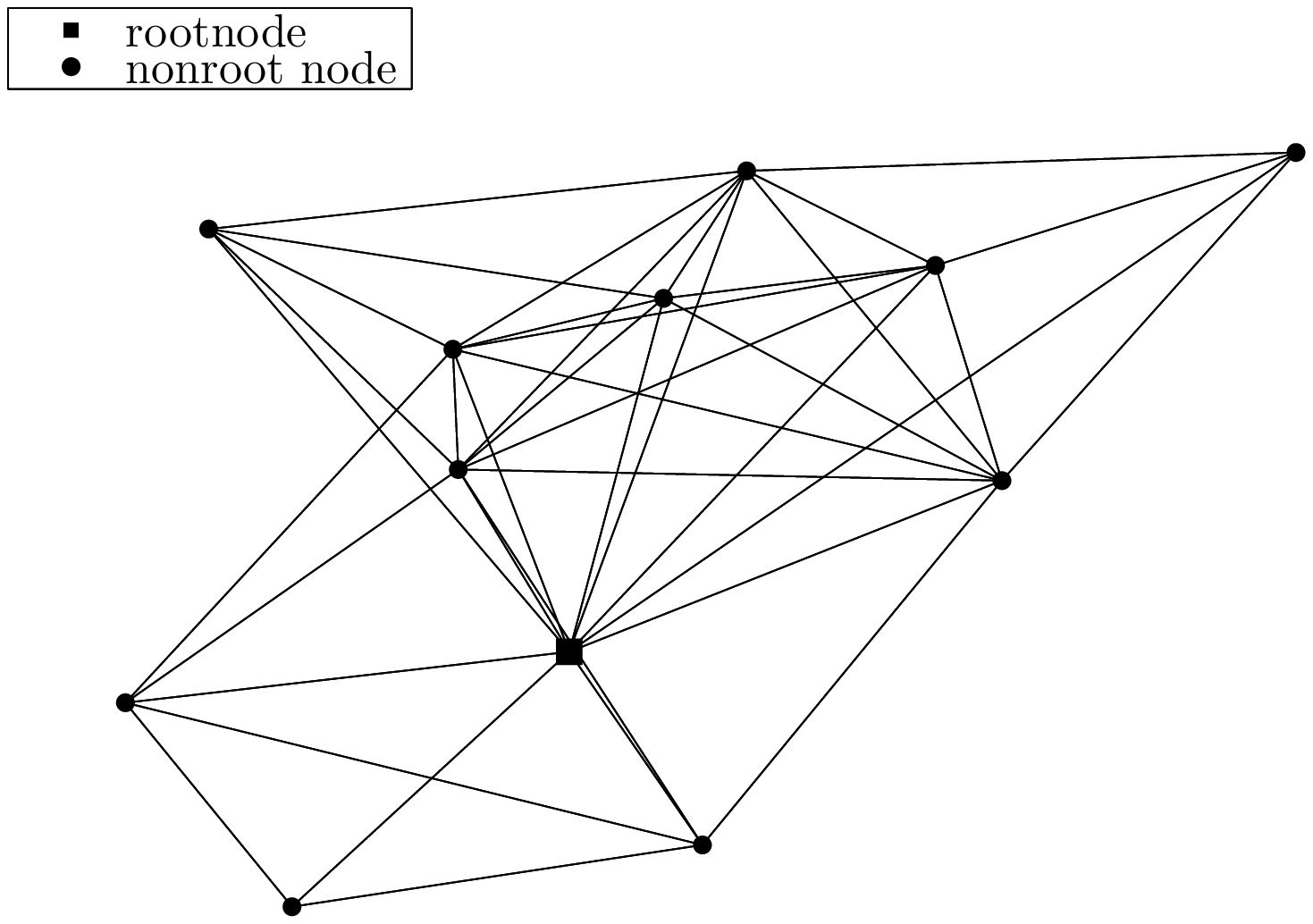}} 
\subfigure[$W_p=10$, $W_l=10$,]{\includegraphics[width = 0.55\columnwidth]{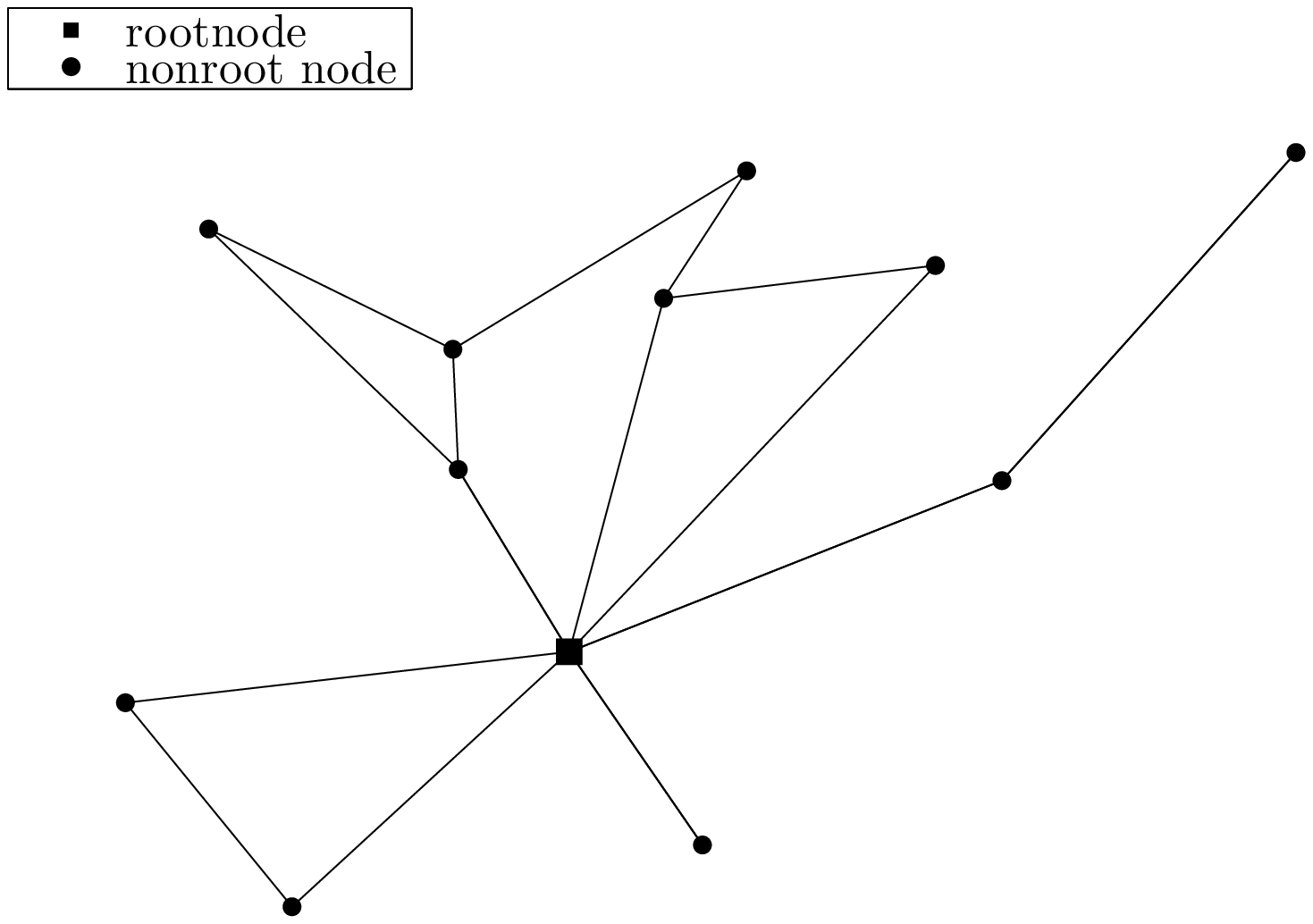}}
\caption{Optimized network topology for different cost models. Different network topologies result in different power-link usage trade-offs.} 
\label{FD_Cellular_fig_toplogycomparison}
\end{figure*}

In Fig.~\ref{FD_Cellular_fig_toplogycomparison}, the optimized network topology is depicted for different cost models for a network with $\mathbb{M}=1$ and $\mathbb{R}=1$. It is observed that different network topologies result in a different power-link usage trade-off. As expected, the strategy with the focus on the cost of power prefers the establishment of a higher number of links, compared to the star topology which provides connectivity through the network with the minimum number of the wireless links. 

\begin{figure*}[h]  
\hspace{1.0cm} \subfigure[channel re-tunning]{\includegraphics[angle = 0, width = 0.60\columnwidth]{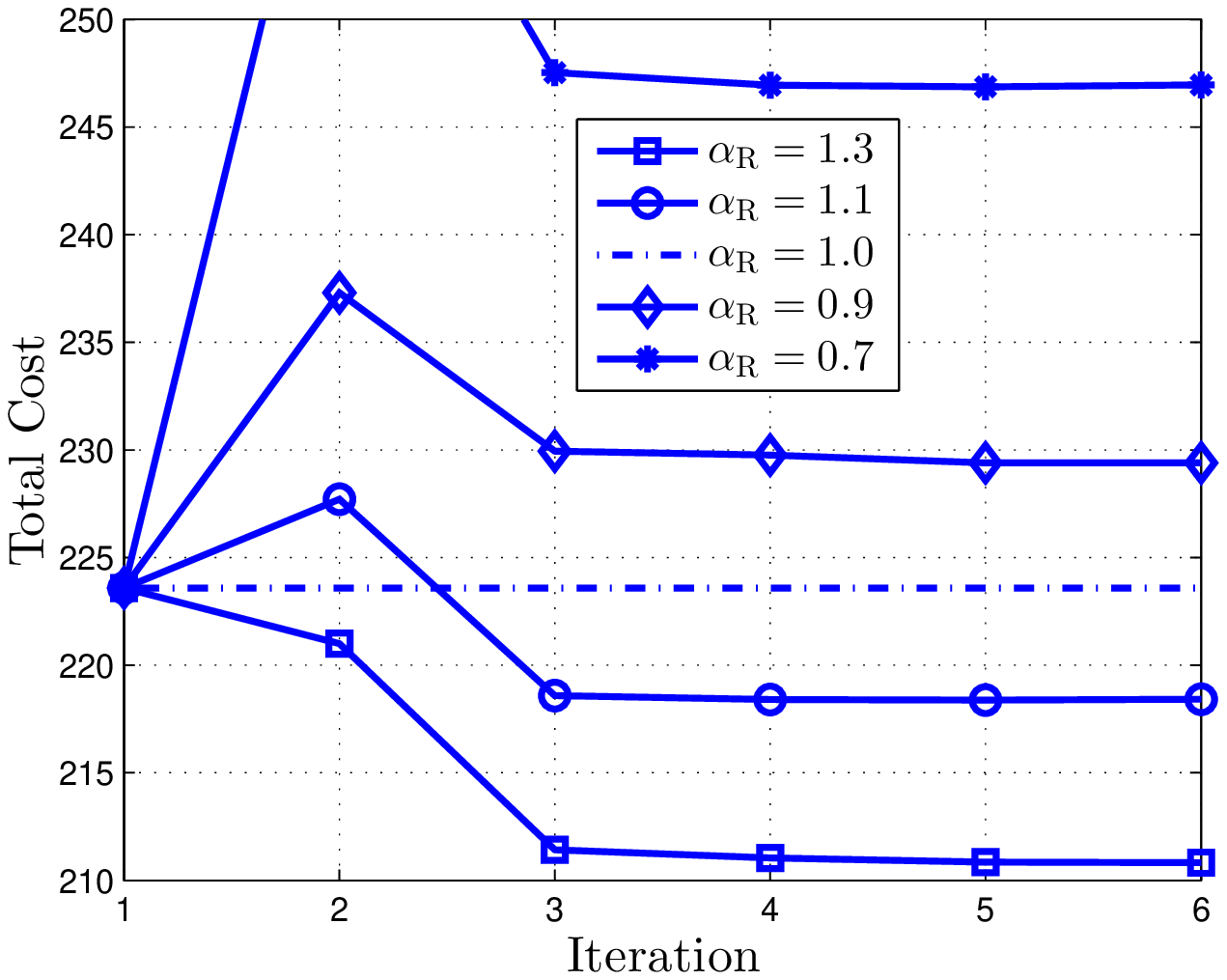}} 
\subfigure[noise re-tunning]{\includegraphics[width = 0.60\columnwidth]{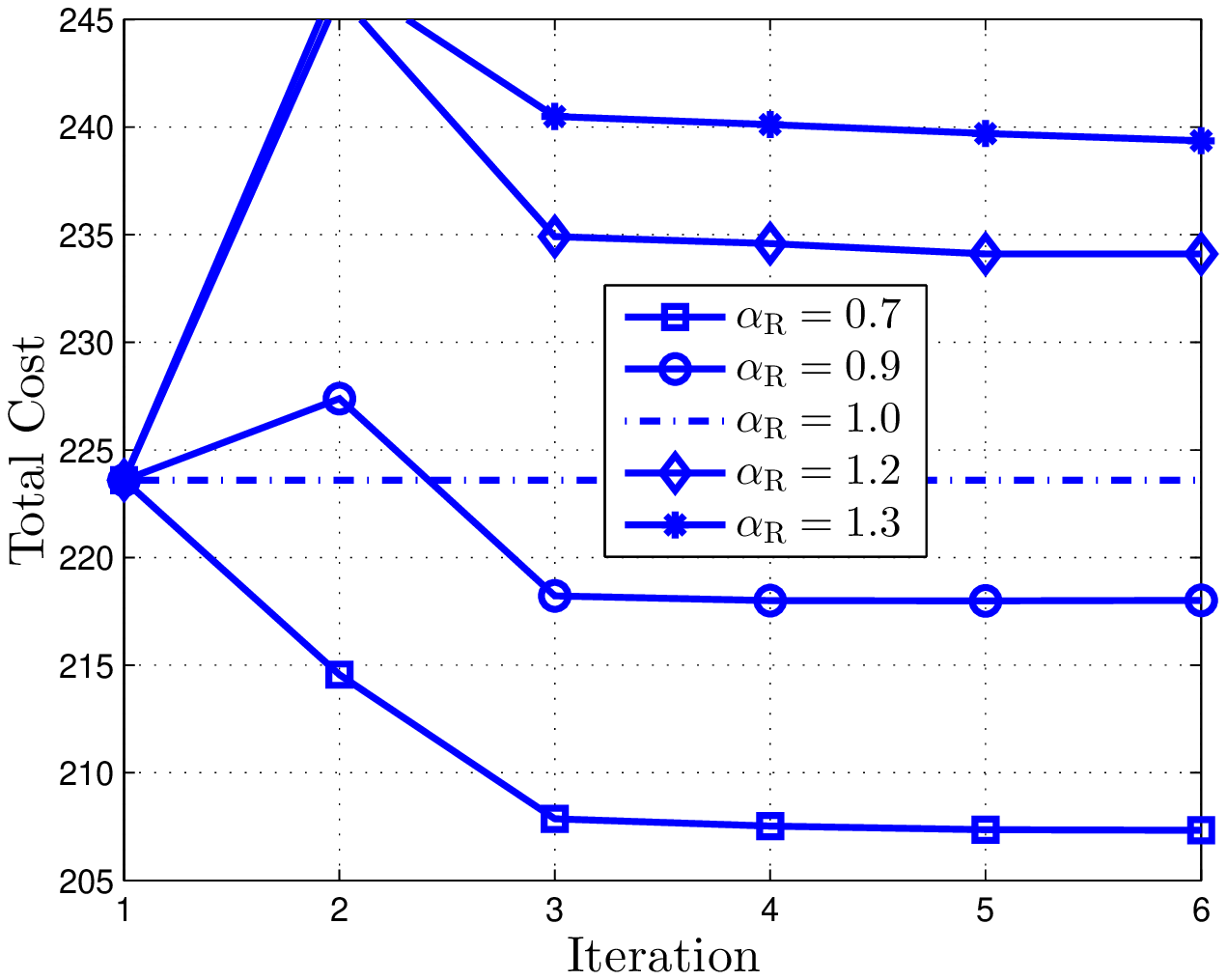}} 
\subfigure[rate re-tunning]{\includegraphics[width = 0.60\columnwidth]{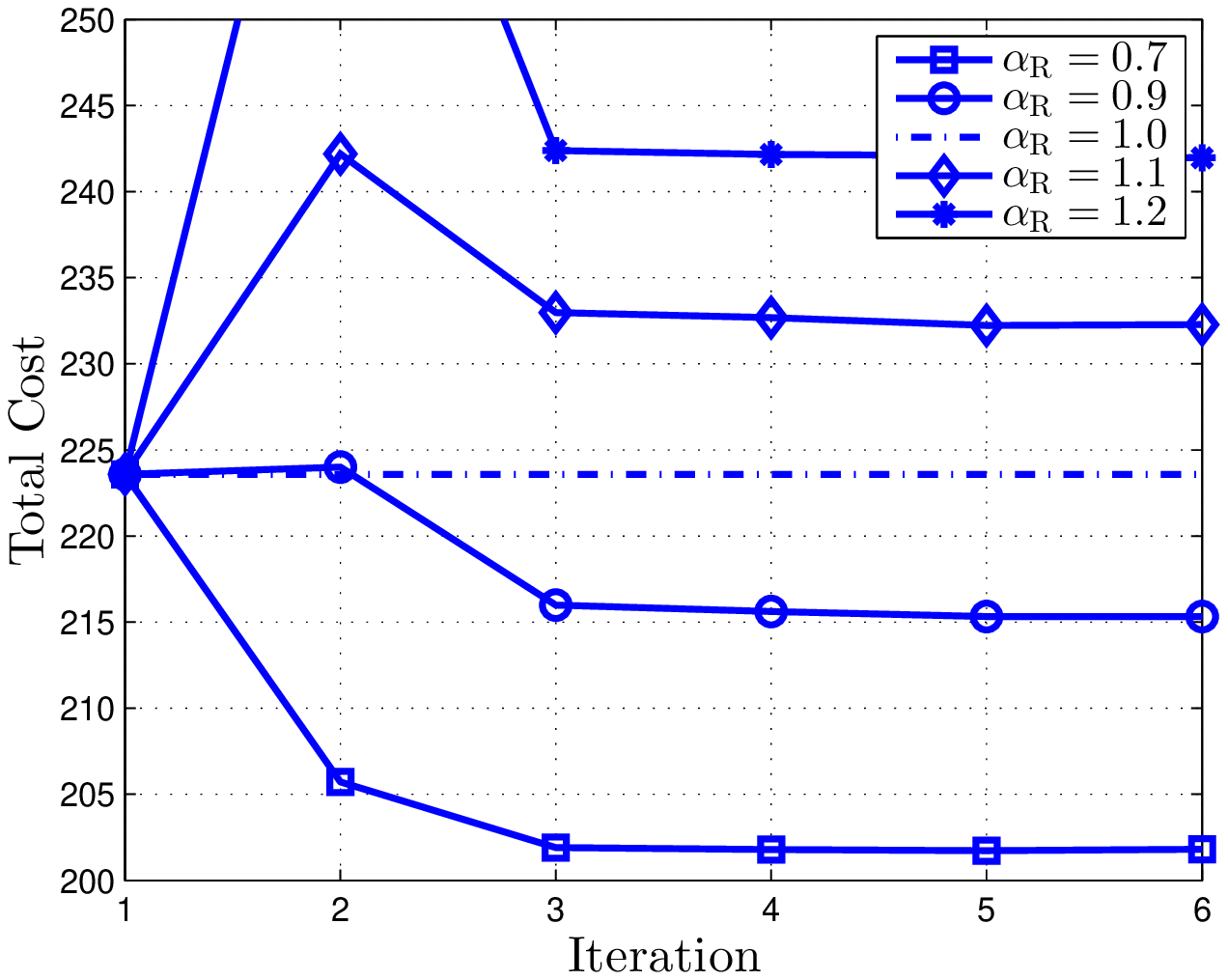}} 
\caption{Iterative network re-tuning. The network reacts to the small changes in the rate requirements and the channel and noise conditions by adjusting the transmit power at each link.} 
\label{FD_Cellular_fig_Retunning}
\end{figure*}

The performance of the proposed iterative network re-tuning is depicted in Fig.~\ref{FD_Cellular_fig_Retunning}~(a)-(c), for different levels of fluctuations in the channel, noise, and rate requirements. The algorithm starts with the outcome of the planning given in Algorithm~\ref{FD_Cellular_alg_1} as the initial point, however, iteratively adjusts the transmit power to comply with the new network conditions. In particular, the ratio $\alpha_{\text{R}}$ indicates the level by which the specific parameters is scaled. Note that due to the infeasibility of the given initial point from the Algorithm~\ref{FD_Cellular_alg_1}, as a result of the scaled parameters, the network power consumption may raise at the initial re-tunning iteration. Nevertheless, the network power consumption is decreased monotonically after the second iteration, and converges in $3$-$6$ iterations. Please note that the benefits of enabling the proposed retunning method is twofold. Firstly, it obtains a feasible network operation point, adjusting to the new rate requirements, noise, or channel conditions. Secondly, it enables the network to save energy, by opportunistically benefiting from the reduced rate demand, when network traffic load is not high. 

In Table~\ref{FD_Cellular_tab_CPUTime} the average per-iteration CPU time is reported for both Algorithms~\ref{FD_Cellular_alg_1} and~\ref{FD_Cellular_alg_1}, for different node cluster sizes\footnote{The simulations are performed on a Linux Debian system with processor Intel Core i$7-3770$S CPU @$3.10$GHz X 4, RAM of $8$GB. The version $2016$b of MATLAB was used along with CVX $2.1$ and Gurobi $7.0$ Solver}. It is observed that a larger problem dimension, i.e, $d:=|\mathbb{L}|\times |\mathbb{F}|$, results in a higher computational time for both algorithms, however, remains below $10$ Hrs for $d \leq 1200$ on a standard user processor. Moreover, due to the proposed SIA framework, the Algorithm\ref{FD_Cellular_alg_2} results in a significantly smaller computational load, due to the efficient convex problem structure. 

\begin{table}[!h]
	\centering
	\caption{Required CPU Time}	
	\label{FD_Cellular_tab_CPUTime}
	\begin{tabular}{|c|c|c|c|c|}
		\hline
		$\vert \mathbb{N}\vert$  & $\vert \mathbb{L} \vert$ & $\vert\mathbb{F}\vert$ & Problem Size & Execution Time in seconds\\
	%	& $w_p = 1, w_f = 20, w_l = 10 $ & $w_p = 1, w_f = 20, w_l = 10 $ & $w_p = 1, w_f = 20, w_l = 10 $ \\
		\hline
		8  & 12  & 6     & 72    & $\begin{array}{l}  360~\text{(MILP)}\\ 1.2~\text{ (re-tunning)}\end{array} $ \\
 		%15 & 24  & 6     & 144   & 854 s\\   
		15 & 24  & 6     & 144   & $\begin{array}{l}   1020 ~\text{(MILP)}\\ 5.6~\text{ (re-tunning)}\end{array} $ \\  
		%15 & 30  & 8     & 240   & 2165 s\\ 
		23 & 40  & 10    & 400   & $\begin{array}{l}   10200~\text{(MILP)}\\   20.1~\text{ (re-tunning)}\end{array} $ \\ 
		%23 & 100 & 10    & 1000  & 14405 s\\ 
		30 & 120 & 10    & 1200  & $\begin{array}{l}   30050~\text{(MILP)}\\   262.8~\text{ (re-tunning)}\end{array} $	\\							
		\hline
	\end{tabular}
\end{table}

\section {Conclusion} \label{sec_conclusion}
With almost all of the below $6$ GHz spectrum already assigned, and the $1000$-fold expected increase of data traffic over the next decade, the need for spectral efficient solutions is apparent in the context of cellular wireless communication systems. In this work, application of FD wireless links is studied as a spectrum-saving mechanism for the wireless backhaul networks. In particular, the coexistence of multiple wireless links at the same channel resource is investigated, utilizing an environment-aware interference management scheme, leading to a reduced overall cost. Moreover, a reactive network re-tuning method is proposed which reacts to small changes in the network data, e.g., QoS requirements or channel conditions, via transmit power adjustment on each wireless link. Numerical simulations suggest that for a dense urban deployment, the proposed methodologies result in the reduction of up to $20\%$ in the overall network cost compared to the half-duplex counterparts.

%\section{Acknowledgement} 
%This work was partly supported by the Deutsche Forschungsgemeinschaft (DFG) grant MA 1184/34-1 (DupLiNk). 

%\appendix
%\input{./Sections/main_appendix}

%%%%{
%%%%%\ifthenelse{\boolean{publ}}{\footnotesize}{\small}
 %%%%\bibliographystyle{IEEEtran}  % Style BST file
%%%%\bibliography{G:/Documents/MyPapers/REFERENCES/references}
 %%%%
%%%%
%%%%}    

% Generated by IEEEtran.bst, version: 1.14 (2015/08/26)

% that's all folks

\end{document}